\documentclass[twocolumn]{aastex63}
\usepackage[utf8]{inputenc}%
\catcode`\@=11
\newcommand{\gsim}{${\mathrel{\mathpalette\@versim>}}$}
\newcommand{\lsim}{${\mathrel{\mathpalette\@versim<}}$}
\newcommand{\@versim}[2]{\lower 2.9truept \vbox{\baselineskip 0pt \lineskip
    0.5truept \ialign{$\m@th#1\hfil##\hfil$\crcr#2\crcr\sim\crcr}}}
\catcode`\@=12

\newcommand{\cmthree}{cm$^{-3}$}

\newcommand{\ddeg}{$^{o}$}
\newcommand{\kms}{km s$^{-1}$}
\newcommand\HI{H\,{\sc i}}
\newcommand\HII{H\,{\sc ii}}
\newcommand\CI{C\,{\sc i}}
\newcommand\CII{C\,{\sc ii}}
\def\be{\begin{equation}}
\def\bea{\begin{eqnarray}}
\def\ee{\end{equation}}
\def\eea{\end{eqnarray}}

\begin{document}
\title{Arecibo-Green Bank-LOFAR Carbon Radio Recombination Line observations toward cold \HI\ Clouds}
\author{D. Anish Roshi}
\affiliation{Arecibo Observatory/University of Central Florida}
\author[0000-0002-5187-7107]{W. M. Peters}
\affiliation{Naval Research Laboratory, Washington, DC}
\author{K. L. Emig}
\altaffiliation{Jansky Fellow of the National Radio Astronomy Observatory}
\affiliation{National Radio Astronomy Observatory,  520 Edgemont Rd, Charlottesville, VA 22903, USA}
\author{P. Salas}
\affiliation{Green Bank Observatory, 155 Observatory Road, Green Bank, WV 24915, USA}
\author{J. B. R. Oonk}
\affiliation{Leiden Observatory, Leiden University, P.O. Box 9513, NL-2300 RA Leiden, The Netherlands}
\affiliation{Netherlands Institute for Radio Astronomy (ASTRON), Postbus 2, 7990 AA Dwingeloo, The Netherlands}
\affiliation{SURF/SURFsara, Science Park 140, 1098 XG Amsterdam, The Netherlands}
\author{M. E. Lebr\'on}
\affiliation{Department of Physical Sciences, University of Puerto Rico, R\'{i}o Piedras Campus, San Juan, PR 00931, USA}
\author{J. M. Dickey}
\affiliation{School of Natural Sciences, University of Tasmania, Hobart, TAS 7001, Australia}

\begin{abstract}
We present results from a search for radio recombination lines in three \HI\ self-absorbing (HISA) clouds at 750 MHz and 321 MHz with the Robert C. Byrd Green Bank Telescope (GBT), and in three Galactic Plane positions at 327 MHz with the Arecibo Telescope. We detect Carbon Recombination Lines (CRRLs) in the direction of DR4 and DR21, as well as in the galactic plane position G34.94+0.0. We additionally detect Hydrogen Recombination Lines (HRRLs) in emission in five of the six sightlines, and a Helium line at 750 MHz towards DR21. Combining our new data with 150 MHz LOFAR detections of CRRL absorption towards DR4 and DR21, we estimate the electron densities of the line forming regions by modeling the line width as a function of frequency. The estimated densities are in the range 1.4 $\rightarrow$ 6.5 cm$^{-3}$ towards DR4, for electron temperatures 200 $\rightarrow$ 20 K. A dual line forming region with densities between 3.5 $\rightarrow$ 24 cm$^{-3}$ and 0.008 $\rightarrow$ 0.3 \cmthree\ could plausibly explain the observed line width as a function of frequency on the DR21 sightline. The central velocities of the CRRLs compare well with CO emission and HISA lines in these directions. The cloud densities estimated from the CO lines are smaller (at least a factor of 5) than those of the CRRL forming regions. It is likely that the CRRL forming and \HI\ self-absorbing gas is located in a denser, shocked region either at the boundary of or within the CO emitting cloud. 

\end{abstract}


\section{Introduction}
Stars form from the collapse of molecular clouds, which must, in turn, be formed continuously in order to sustain star formation in the Galaxy. It has been suggested that a fraction of the interstellar regions exhibiting \HI\ self-absorption (HISA) are sites of molecular cloud formation \citep{Gibson2000,Klaassen2005}. HISA occurs when foreground cold atomic gas absorbs the 21cm emission line from warm \HI\ gas behind it. Many HISA regions have a low atomic fraction but little or no $^{12}$CO emission, implying that a significant amount of ``CO dark'' molecular gas is present \citep{Klaassen2005}. In other cases the HISA is well mixed with CO emitting molecular gas \citep{Li2003}.

While spectral line studies have inferred molecular gas content in HISA regions, observational constraints on the molecular formation process are still lacking. There are a few possible molecule formation processes that have been discussed in the literature \citep[see review by][]{Dobbs2014}. These include a converging flow, such as a gas flow caused by the expansion of an \HII\ region or supernova remnant, which could enhance the local gas density 
resulting in the formation of molecules. Another possibility is a quasi-static cloud contraction process where a slight increase in ambient pressure accelerates the phase transition from atomic to molecular gas.

Observations of radio recombination lines (RRLs) of ionized carbon, 
\HI\ 21cm absorption, and CO emission may be used to put constraints on the molecular formation processes in HISA regions. RRL optical depth has a strong inverse dependence on the electron temperature ($\propto T_e^{-2.5}$, where $T_e$ is the electron temperature). Thus RRLs from ionized carbon (CRRLs) are preferentially detected toward cold regions. 
The inferred neutral gas densities toward HISA regions are a few times 10$^3$~cm$^{-3}$ \citep{Li2003,Klaassen2005}. CRRLs from neutral regions with such densities are expected to be brighter at lower frequencies \citep[\lsim 1 GHz;][]{Roshi2011,Kantharia2001,Salgado2017b}.

The detection of low-frequency CRRLs from a well known HISA region -- the Heeschen-Riegel-Crutcher region \citep[HRC,][]{Heeschen1955,Riegel1972} -- was reported by \citet{Roshi2011, Oonk2019}. By combining multi-frequency CRRL and \HI\ data, \citet{Roshi2011} were able to constrain the physical properties and processes in the HRC cloud. These include electron temperature, electron density, line of sight (LOS) path length, hydrogen nuclear density, hydrogen atomic density, and hydrogen molecular density. The derived physical properties along with constraints on the background ultraviolet (UV) radiation field were then used to estimate the H$_2$ formation and dissociation rates in the cloud. They show that the H$_2$ formation rate exceeds the H$_2$ dissociation rate, suggesting that the cloud is in the process of converting \HI\ to H$_2$ and will convert all of its atomic hydrogen into the molecular form over a time scale \gsim$10^{5}$ years.

In this paper, we expand our observational study toward the HISA regions { detected along the sightlines toward the \HII\ region DR21, and the supernovae remnants (SNRs) DR4 \& HB21.} These  were selected based on HISA detection in the Leiden/Argentine/Bonn 21cm line survey data \citep{Kalberla2005}. Unlike the HRC cloud, the selected directions 
have intense background Far-UV radiation fields due to the presence of star forming regions. These observations therefore will allow us to probe molecule formation in very different environments compared to the HRC cloud. Radio recombination line observations were made with the GBT at 321 and 750 MHz.  We also present results from blind recombination line observations made toward three positions (G34.20+0.0, G34.94+0.0, G35.17+0.0) in the Galactic plane with the Arecibo telescope. These observations were made at 327 MHz. The positions were selected because an earlier CRRL survey has detected lines in this part of the galactic plane with an angular resolution of 2\ddeg $\times$ 0.5\ddeg\citep{Roshi2002}. The observations and data analysis procedure are described in Section~\ref{sec:obs} and \ref{sec:datanalysis} respectively. The results of recombination line observations and the estimation of physical properties by combining our data with { LOw Frequency ARray \citep[LOFAR;][]{vanHaarlem2013} and other previous} CRRL observations (Oonk et al. in preparation) and existing \HI\ and CO line data are presented in Section~\ref{sec:result}. A summary of the results and future work are given in  Section~\ref{sec:sum}. 

\section{Observations}
\label{sec:obs}

The source name, coordinates, date of observations, center frequency of observations, full-width at half maximum (FWHM) beamwidth and the telescope used for the observations are given in Table~\ref{tab:obssum}. 


The 321 and 750 MHz observations with the GBT (Project code GBT12A-352) used the Green Bank Spectrometer for measuring the power spectrum. For the 750 MHz observations, the spectrometer was configured to have eight 12.5 MHz sub-bands, with each sub-band divided into 4096 spectral channels. The sub-bands were tuned to observe eight $\alpha$ RRL transitions (n=201, 203, 205 to 210, $\Delta n =1$) within the PF1\_800 receiver bandwidth.
We use the quantum number n=206 when referring to the final spectrum at this frequency obtained by averaging all the transitions (see Section~\ref{sec:datanalysis}; also Table~\ref{tab:rrlpar}). The RRL transitions were selected such that they are located at frequency ranges relatively free of strong radio frequency interference (RFI). We observed in `standard frequency switch' mode, with a switching period of 1 sec and frequency offset of 2.0 MHz ($\sim$ 815 \kms\ at the observing frequency). 

For the 321 MHz observations, we used three 12.5 MHz sub-bands with 8192 spectral channels each to observe RRLs with the PF1\_342 receiver.  The 3 sub-bands were centered at the observing frequencies 325.15, 309.62 and 341.87 MHz. The separation between RRL transitions near 340 MHz is $\sim$ 3.7 MHz. Thus multiple RRL transitions were observed in each sub-band. The RRLs observed are the $\alpha$ transitions corresponding to n = 268, 271, 273, 275, 276 and 277. We use the quantum number n=273 when referring to the final spectrum at this frequency obtained by averaging all the transitions (see Section~\ref{sec:datanalysis}; also Table~\ref{tab:rrlpar}).  We observed in `standard frequency switch' mode, with a switching period of 1 sec and frequency offset of 1.0 MHz ($\sim$ 910 \kms). We used 9-level sampling mode since observations were to be made in the presence of RFI, although these transitions were relatively free of strong RFI. We observed 3C 295 to determine the telescope pointing correction and for flux density calibration. 

We used the Mock spectrometer for the 327 MHz Arecibo observations (Project code A3333). The 327 MHz receiver has a 3dB bandwidth of $\sim$ 30 MHz, and there are eight $\alpha$ RRL transitions (principal quantum number n=268 to 275, $\Delta n = 1$) within this frequency range.  We configured the Mock spectrometer with a 160 MHz clock frequency and set the bandwidth divisor to 4 to observe all eight RRL transitions simultaneously. This configuration provided 40 MHz observing bandwidth with 8192 spectral channels, for a spectral resolution of 4.9 kHz. The observations were made in a `pseudo' frequency switching mode based on the standing waves in the blocked aperture of the telescope, which have a separation in frequency of $\sim 1.1$ MHz.  Each one minute scan in 'standard on' mode was followed by scans offset by $-500$ and $+500$ \kms from the source velocity.  At the central rest frequency of 326.5 MHz, this corresponds to switching the first local oscillator frequency by $\sim 545$ kHz. The $\pm 500$ \kms\ shift allowed us to observe all eight hydrogen and carbon RRLs simultaneously, because they are separated by only 149.6 \kms. The power spectrum was integrated for 1 sec, and 60 of them were written to a fits file every minute at the end of each scan.  We observed a `standard ON/OFF' scan with calibrated noise switching at the beginning of each observation to measure the on-source continuum antenna temperature. We also observed a flux density calibrator, 3C 394, in `standard ON/OFF' mode with noise switching to calibrate the noise cal value in Jy.

The power spectrum was integrated for 1 sec, and 60 of them were written to a fits file every minute at the end of each scan.  We observed a `standard ON/OFF' scan with calibrated noise switching at the beginning of each observation to measure the on-source continuum antenna temperature. We also observed a flux density calibrator, 3C 394, in `standard ON/OFF' mode with noise switching to calibrate the noise cal value in Jy.

\section{Data reduction}
\label{sec:datanalysis}


We used GBTIDL\footnote{\url{https://gbtidl.nrao.edu/}}\footnote{\url{https://www.l3harrisgeospatial.com/Software-Technology/IDL}} to create 1 second integration bandpass corrected spectra for the GBT observations. GBTIDL routines were also used to correct the bandpass for each sub-band from the frequency switched data set. The GBTIDL routine vshift was used to apply LSR velocity correction for each CRRL transition. The bandpass and velocity corrected spectra were written to a text file for each CRRL transition. These spectra were then edited for RFI.  

The RFI affected data may be sorted into three types: (a) A majority (90\%) of the RFI were narrow band, confined to one or two spectral channels. The amplitude of each varies slowly over several tens of minutes. (b) In a few cases (40\% of the observing time) the frequency and strength of the RFI were time dependent, but could be easily identified in 1 sec integrated spectra. (c) Rarely (10\% of the observing time), the frequency of RFI sweeps across a large fraction of the sub-bands in a few seconds. Excising the RFI thus requires careful editing of the data both in time and frequency domains, for which we developed an RFI editing program in MATLAB\footnote{\url{https://www.mathworks.com/products/matlab.html}}. 

The RFI editing program uses a channel weighting scheme to edit out the RFI. Each channel is assigned unity weight at the start. If a channel is identified as affected by RFI or corrupted by instrumental problems, then the weight of that channel is set to zero. The weighted spectra are then averaged across time. Identification of RFI is currently done manually by examining each 1 sec spectrum.  For the GBT data sets, the weighted average spectrum forms the final spectrum for each RRL transition and sight line. 

\begin{figure*}
\includegraphics{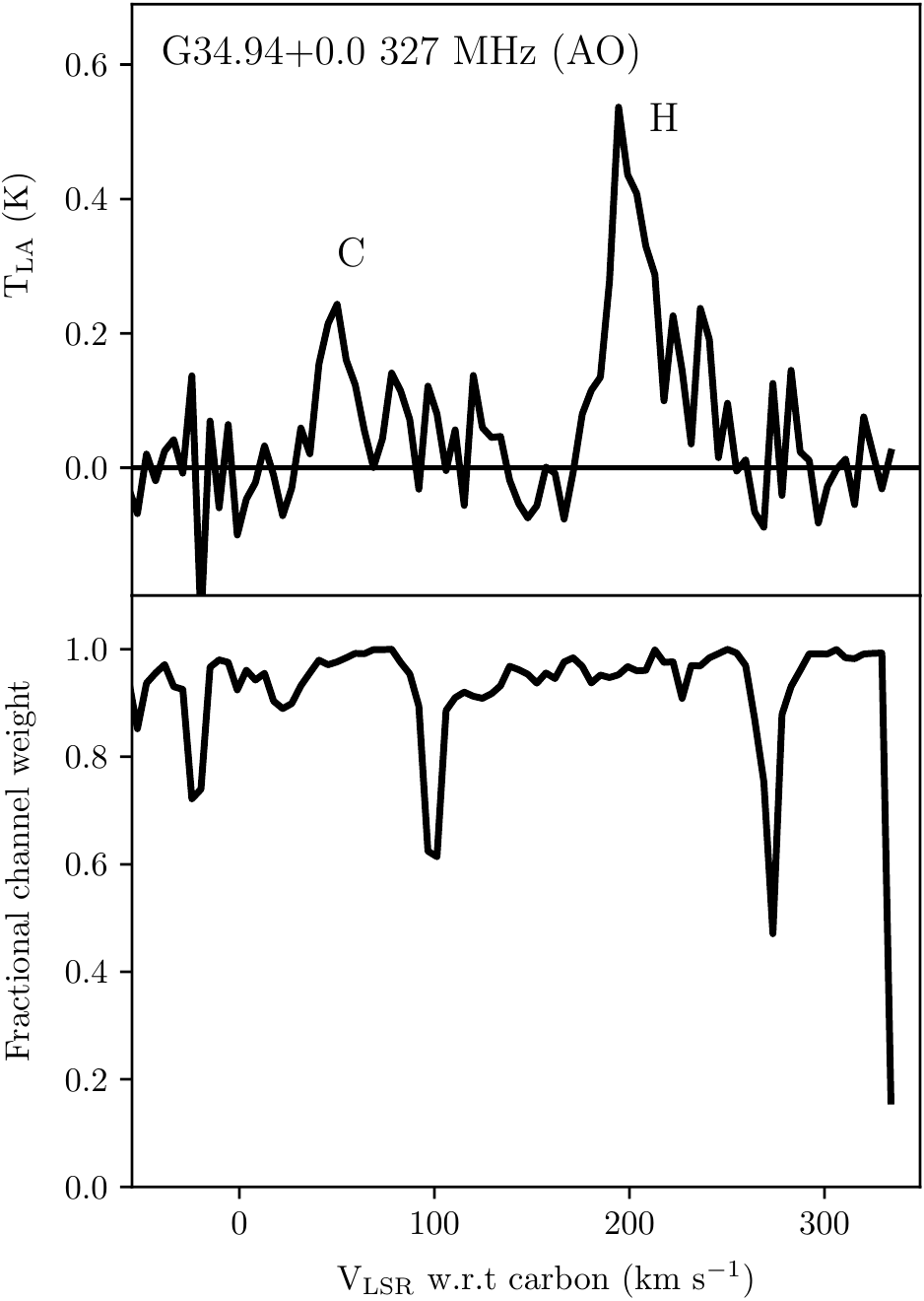}
\begin{center}
\caption{Recombination line spectrum along with the fractional channel weights. The weighted averaged spectrum toward G34.94+0.0 obtained with the Arecibo telescope at 327 MHz is shown on the top plot and the fractional channel weights are shown on the bottom plot. The typical variation in weights across channels in the final averaged spectrum is about 10\%. } 
\label{fig:rrlwt} 
\end{center}
\end{figure*}

We followed similar steps to analyze the Arecibo data set. Phil Perillat's IDL routines\footnote{\url{http://www.naic.edu/~phil/software/software.html}} were used to generate a spectrum covering the full 40 MHz bandwidth. 
A spectrum covering a 1000 \kms\ range centered on each of the 8 RRL transitions was extracted. 
These spectra were then edited for RFI using the MATLAB RFI-editing routine described above, and the weighted average was taken to get the spectrum corresponding to each RRL transition. Because the observations were made with in-band frequency switching, each RRL spectrum was `folded' to improve the integration time. The spectra were re-sampled using a Fourier transform method \citep{Roshietal2005} and aligned in velocity range. The `folded' spectra for the different transitions were then re-sampled to a common velocity resolution and averaged to get the final integrated spectrum. This analysis method was tested by observing an off-source position for 2 hours with the Arecibo telescope at 327 MHz. No spurious line like feature was detected in this data set up to a 50 mK level. The weights were processed in the same way as the spectrum during the resampling process, so the final spectrum has associated channel weights. These indicate the number of 1 second data points included in each spectral channel. Fig.~\ref{fig:rrlwt} shows an example RRL spectrum toward G34.94+0.0 obtained with the Arecibo telescope along with the weights. The typical variation in weights across the final averaged spectrum is $\sim$10\%. 

\section{Results}
\label{sec:result}

\begin{figure*}
\includegraphics[width=\textwidth]{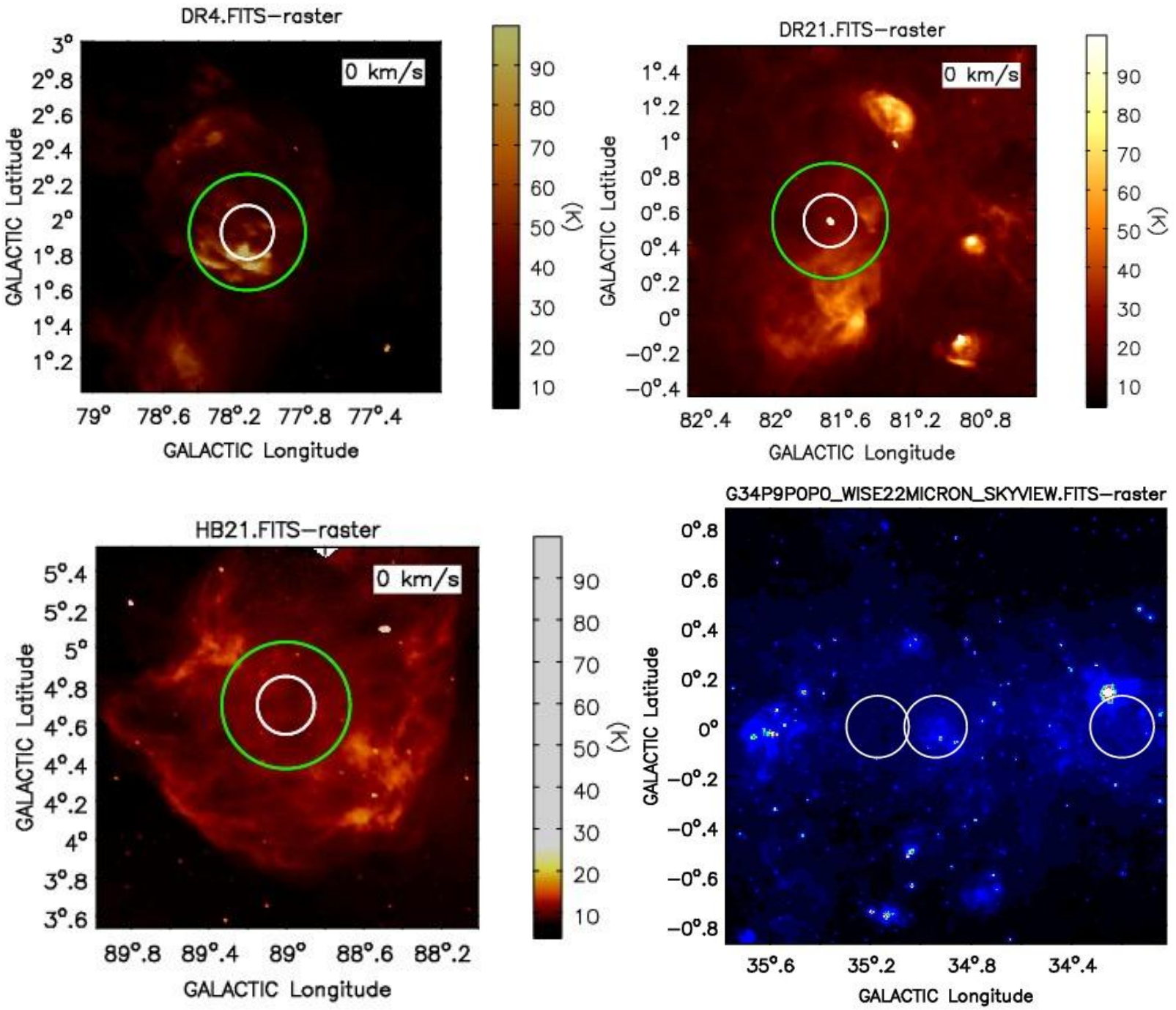}
\begin{center}
\caption{\textbf{Top-left:} The 1.4 GHz CGPS continuum image of DR4. Recombination line observations toward DR4 was made with the GBT. The green and white circles represent the FWHM beam width at 321 MHz ($\sim$0.66\ddeg) and 750 MHz respectively( $\sim$0.3 \ddeg).  \textbf{Top-right:} Same as top-left figure but toward DR21. \textbf{Bottom-left:} Same as top-left figure but toward HB21. \textbf{Bottom-right:} The three positions observed with the Arecibo telescope near 327 MHz are marked on the {22$\mu$m} WISE image of the galactic plane. The white circles represent the FWMH beam width at 327 MHz ($\sim$ 0.25\ddeg).} 
\label{fig:obspos} 
\end{center}
\end{figure*}

\begin{figure*}
\begin{center}
\includegraphics{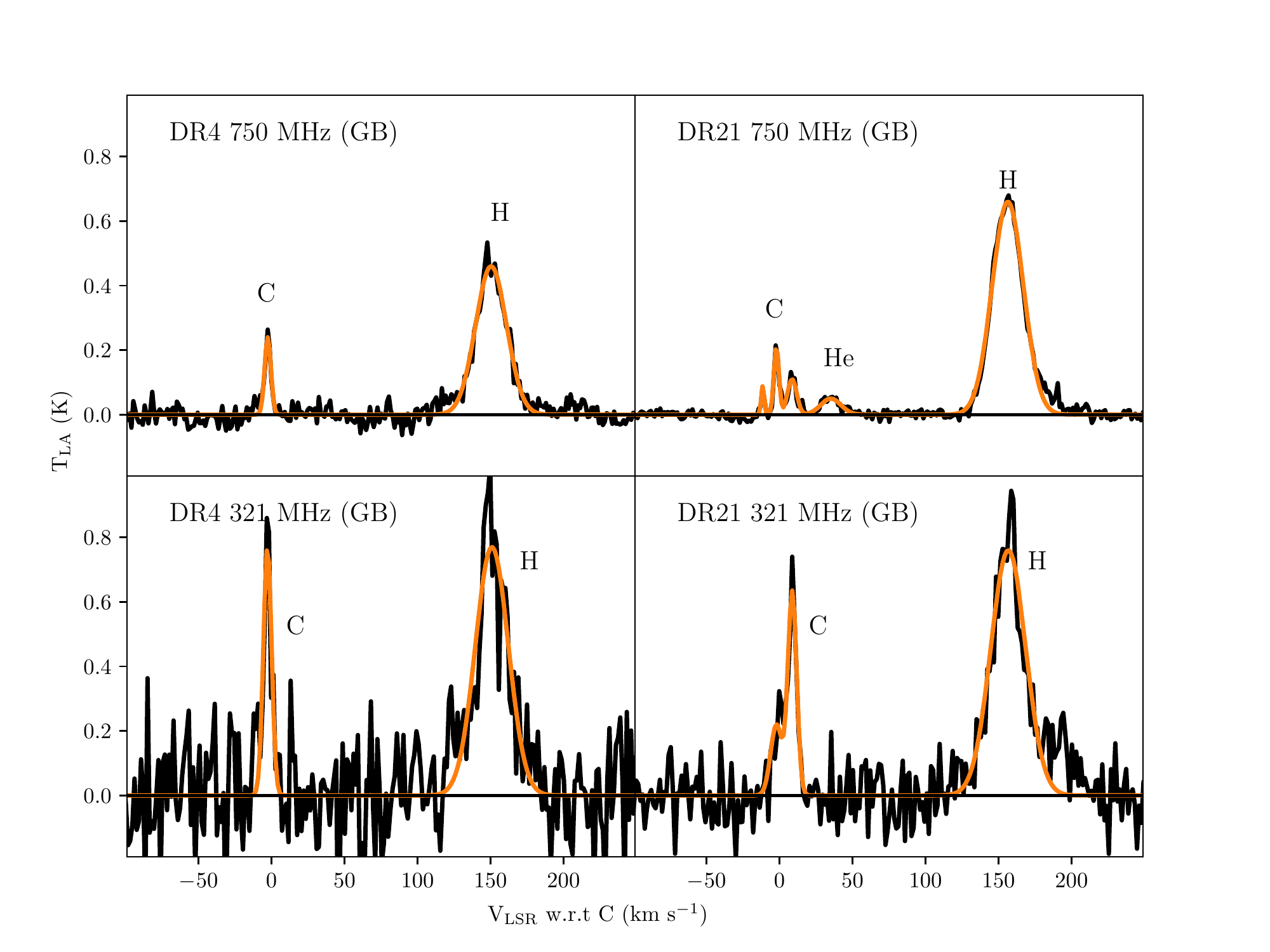}
\caption{Radio recombination line spectra along with Gaussian line models toward DR4 and DR21. The data toward the two sources were obtained using the GBT at 750 and 321 MHz. } 
\label{fig:spec1} 
\end{center}
\end{figure*}

\begin{figure*}
\begin{center}
\includegraphics{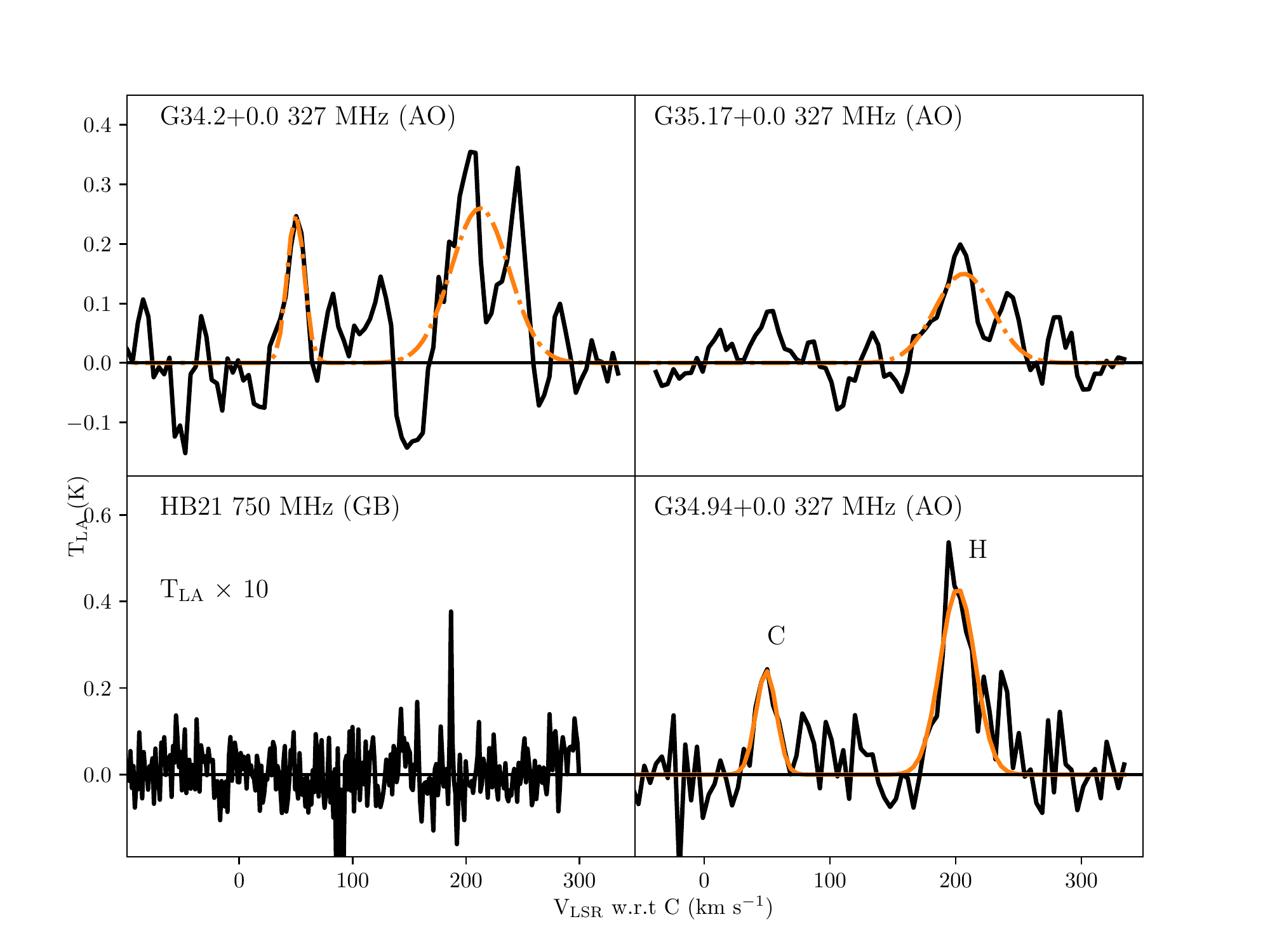}
\caption{Radio recombination line spectra toward G34.2+0.0, G35.17+0.0, G34.94+0.0 and HB21.The data toward the first three sources were obtained with the Arecibo Telescope at 327 MHz and those toward HB21 were obtained with the GBT at 750 MHz. The Gaussian line models (orange solid and dash-dot curves; latter indicates tentative detection) are over plotted in those spectra where lines were detected. } 
\label{fig:spec2} 
\end{center}
\end{figure*}

The observed positions are shown in Fig.~\ref{fig:obspos} (see Table~\ref{tab:obssum}). The RRL spectra obtained toward these positions are shown in Figs.~\ref{fig:spec1} \& \ref{fig:spec2}. The frequency of observation, atom producing the RRL, line antenna temperature, LSR (Local Standard of Rest) velocity and FWHM line width obtained from the spectra are given in Table~\ref{tab:rrlpar}. CRRLs were detected toward DR4, DR21, G34.94+0.0 and G34.2+0.0 (tentative detection) at all of the observed frequencies. Hydrogen lines were detected toward positions DR4, DR21, G34.90+0.0, G34.2+0.0 (tentative detection) and G35.17+0.0 (tentative detection) at all of the observed frequencies. Helium and heavy element
recombination lines were detected towards DR21 at 750 MHz. No lines were detected towards HB21, with a limiting sensitivity of {5 mK}.  The listed line widths are corrected for broadening due to spectral resolution using the equation $\sqrt{\Delta V_{obs}^2 - \Delta V_{res}^2}$, where $\Delta V_{obs}$ is the observed line width and $\Delta V_{res}$ is the spectral resolution.

In the following subsections we combine our. new RRL observations with LOFAR { and other existing} CRRL, \HI\ and CO data sets to constrain the physical properties of the line forming regions. { The implicit simplifying assumptions made in the modeling presented below to estimate the physical properties are that the line forming region is homogeneous gas with uniform temperature and density. } A brief discussion of the observed hydrogen and helium line formation is also presented. { We note here that the LOFAR detections of CRRLs in absorption near 150 MHz towards DR4 (see Section~\ref{subsec:dr4crrl}) and DR21 (see Section~\ref{subsec:dr21crrl}) are the highest frequency carbon recombination line absorption reported to date.}  

\subsection{DR4}
\label{sec:dr4}

DR4 is a shell-type SNR located in the Cygnus-X region. A detailed study of the \HI\ 21cm line associated with the SNR was done by \citet{Landecker1980} and \citet{Ladouceur2008}. They used data from the DRAO synthesis telescope; the latter authors also used the CGPS data set \citep{Taylor2003}. The angular resolution of the 21 cm observations was better than 2\farcm5. The \HI\ line emission and absorption in this direction are quite complex. The 21 cm line analysis indicates that the SNR blast wave is compressing the \HI\ gas into a shell. The velocity structure of the \HI\ line could be explained as an expanding shell (either due to the SNR blast or the stellar wind of the progenitor) with speed $\sim$ 25 \kms. The near-side of the shell is seen in \HI\ absorption against the bright SNR and also against the 21 cm emission from the far-side part of the shell (i.e. as \HI\ self-absorption). 

\subsubsection{Cold gas properties derived using 21cm line data}
\label{subsec:dr4hi}

\begin{figure*}
    \centering
     \includegraphics{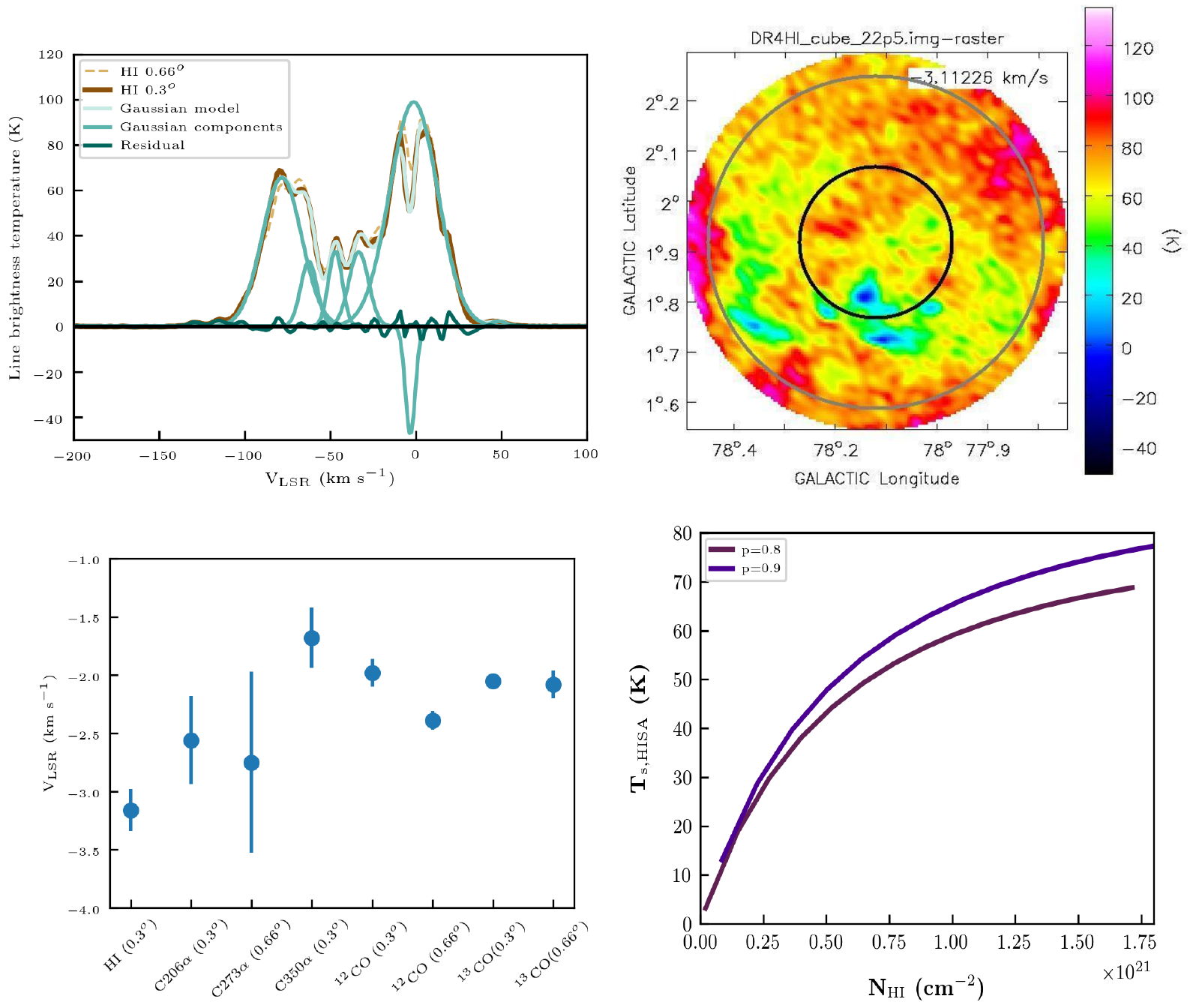}
    \caption{DR4 compilation. \textbf{Top-left:} \HI\ 21cm spectra averaged over 0.66$^{o}$ and 0.3$^{o}$ beams. Gaussian components of the 0.3$^{o}$ averaged \HI\ spectrum are shown along with the net Gaussian line model and residual obtained after removing the line model from the observed spectrum. \textbf{Top-right:} CGPS 21cm image obtained from the spectral channel data near the peak of the \HI\ self-absorption velocity of -3.2 \kms. The angular resolution of the image is $\sim$ 2.5\arcmin. The grey and black circles indicate the GBT beam sizes at 321 MHz and 750 MHz respectively. The peak amplitude of the background emission near -3.2 \kms\ is 99 K. Thus line amplitudes less than 99 K indicate \HI\ absorption. The \HI\ absorption is extended over the GBT beam but the absorption amplitude changes within the beam area. This change could be due to a combination of variations in physical properties and background continuum. \textbf{Bottom-left:} LSR velocities of \HI\ , CRRLs and CO lines observed toward DR4. The error bars shown are $\pm$ 2$\sigma$ values. \textbf{Bottom-right:} The spin temperature and \HI\ column density of the gas responsible for self-absorption for p=0.8 and 0.9 are shown by the solid curves (see Section~\ref{subsec:dr4hi}).}
    \label{fig:DR4HIall}
\end{figure*}

The CGPS \HI\ spectra averaged over an area which matches the GBT beam of 0.66\ddeg\ at 321 MHz and 0.3\ddeg\ at 750 MHz are shown in Fig.~\ref{fig:DR4HIall} (top-left). A Gaussian model for the spectrum extracted at 0.3\ddeg\ resolution is also shown in Fig.~\ref{fig:DR4HIall}. The line amplitudes, central velocities and FWHM line widths of the model components are given in Table~\ref{tab:hicopar}. The central velocity of the \HI\ self-absorption feature agrees within 2$\sigma$ with the central velocity of the 750 and 321 MHz CRRLs (see Fig.~\ref{fig:DR4HIall} bottom-left). Thus we conclude that the \HI\ self-absorption and CRRL forming regions are co-located. 

Fig.~\ref{fig:DR4HIall} (top-right) shows the \HI\ line image at the peak velocity ($\sim$ -3.2 \kms) of the absorption feature. 
The \HI\ absorption is spread over the GBT beam as inferred from the lower line temperature compared to the background emission ($\sim$ 99K; see Table~\ref{tab:hicopar}). The { amplitude of the absorption feature}, however, shows significant variation (see Fig.~\ref{fig:DR4HIall} top-right) which could be due to a combination of variation in background continuum temperature, amplitude of the {background} \HI\ emission and the properties of the cold gas. We conclude that the self-absorption feature is extended over the GBT beam with which the RRL observations were made.

We follow the method outlined by \citet{Denes2018} to constrain the properties of the cold gas responsible for the \HI\ self-absorption. The spin temperature, $T_s$, of the absorbing gas can be written as \citep[Eq. 4 in][]{Denes2018}
\be{
T_s = \frac{T_{ON} - T_{OFF}}{1 - e^{-\tau_{HISA}}} + T_c + p\; T_{OFF}
\label{eq:ts}
}\ee
where $T_{ON}$ and $T_{OFF}$ are respectively the observed \HI\ spectrum and that in the absence of absorption (both obtained after continuum subtraction), $\tau_{HISA}$ is the line optical depth of the absorbing gas,  $T_c$ is the brightness temperature of the continuum emission behind the absorbing gas and $p$ is a parameter used to express the background \HI\ emission as a fraction of $T_{OFF}$.
Eq~\ref{eq:ts} assumes the line optical depths of the background and foreground gases producing the \HI\ emission are negligible. The peak optical depth of the absorbing gas is given by
\be{
\tau_{HISA, peak} = 5.2 \times 10^{-19}\; \frac{N_{HI}}{T_s\;\Delta V}
}\ee
where $N_{HI}$ is the \HI\ column density of the absorbing gas in units of cm$^{-2}$ and $\Delta V$ is the FWHM line width of the \HI\ absorption in \kms.

Gaussian line modeling of the \HI\ spectrum averaged over a 0.3$^{o}$ beam was used to get $T_{ON}$ and $T_{OFF}$, $\Delta V$ of the absorption component (see Table~\ref{tab:hicopar}) and $T_c$ is obtained from the CGPS continuum image.  Fig~\ref{fig:DR4HIall} (bottom-right) shows the cold gas $T_s$ and $N_{HI}$ that are consistent with the observed \HI\ self-absorption for $p = 0.8$ and 0.9; we assume a large fraction of $T_{OFF}$ is located behind the absorbing gas. The continuum emission varies within the GBT beam, 
so we examine the spectra against different bright continuum background sources.   We also assume that the physical properties of the cold \HI\ are similar over the beam area.   
With these assumptions, we estimate that the physical parameters of the cold \HI\ gas are  $T_s < 55 K$ and $N_{HI} < 7 \times 10^{20}$ cm$^{-2}$ for $p = 0.9$.

\subsubsection{Gas properties derived using CRRL data}
\label{subsec:dr4crrl}

\begin{figure*}
\begin{center}
\includegraphics{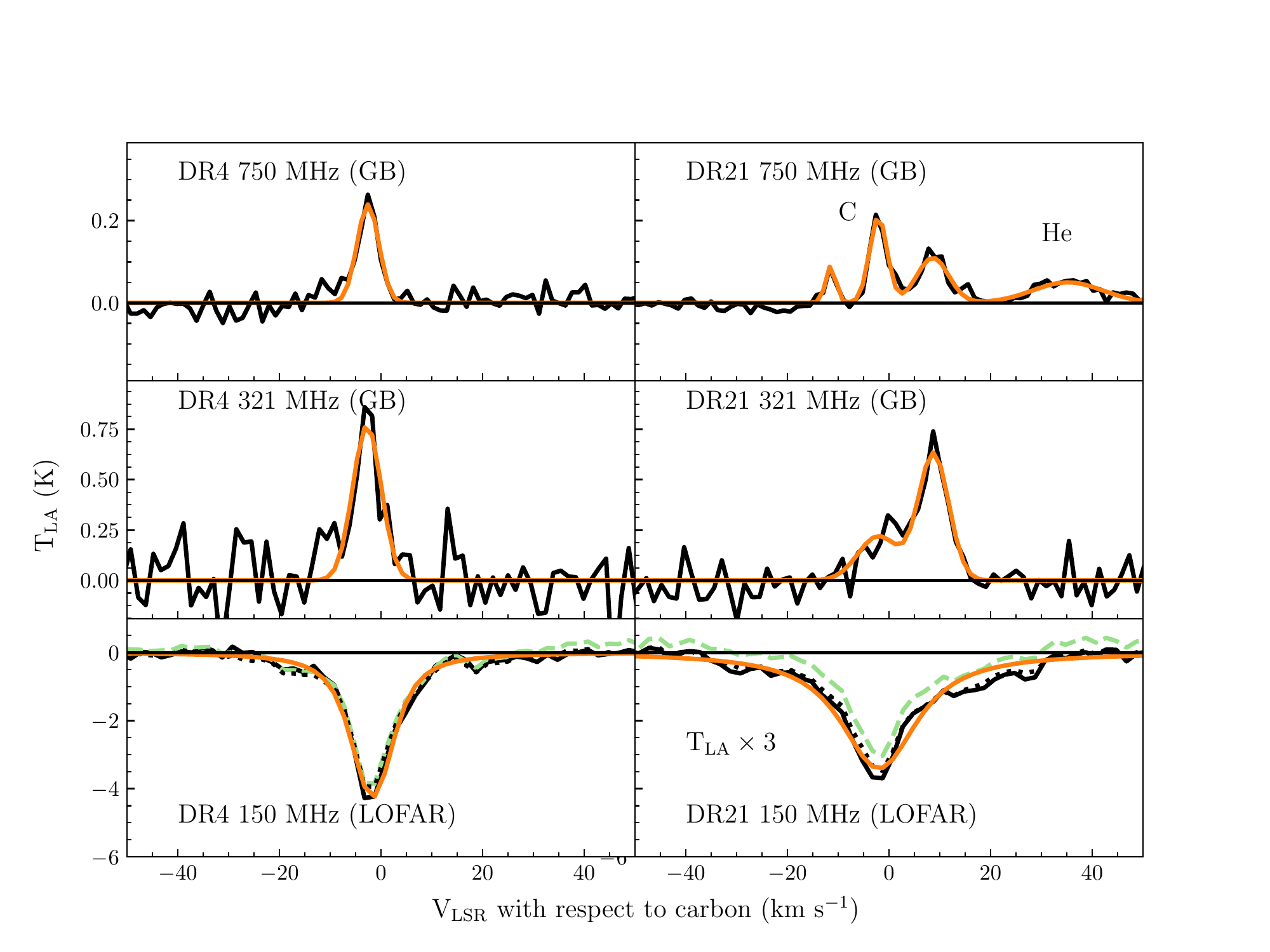}
\caption{Carbon recombination line spectra at 750 (top), 321 (middle) and 150 (bottom) MHz towards DR4 (left) and DR21 (right). The Gaussian line models (orange curves) for the 750 and 327 MHz spectra are over plotted. The 150 MHz spectra averaged over 0.3\ddeg\ region are shown in black solid lines in the bottom panel and the Voigt model fit to these spectra are shown in orange solid curves.  3$^{rd}$ order polynomials were subtracted from these spectra. The 150 MHz spectra averaged over 0.66\ddeg\ region are shown by the dotted and dashed curve. No spectral baselines were subtracted from the spectra shown in dashed curve while 3$^{rd}$ order polynomials were subtracted from the spectra shown in dotted curve.  } 
\label{fig:3crrls} 
\end{center}
\end{figure*}

The carbon line intensity and line width at a given frequency depend on the electron temperature ($T_e$), electron density ($n_e$), LOS path length, background radiation field and the non-LTE parameters characterising the atomic level population \citep[e.g.,][]{Payne1994, Roshi2011, Salas2017, Salgado2017a}. Thus observations of multiple CRRL transitions are required to constrain the physical properties of the gas \citep{Oonk2017}. We combine our data set with the LOFAR CRRL observations toward DR4 at 150 MHz (n=350) to estimate the physical properties. {The angular resolution of LOFAR observations is $\sim$ 10\arcmin\ at this frequency.} The details of the LOFAR observations will be described elsewhere (Oonk et al in prep.). All three CRRLs observed toward DR4 are shown in Fig.~\ref{fig:3crrls}. The LOFAR spectra are obtained by extracting the average line temperature within an aperture equal to the two GBT beam sizes. 

The width of the CRRL is affected by Doppler (thermal and non-thermal), pressure and radiation broadening. The latter two strongly depend on the quantum number, electron density and background radiation field \citep[e.g.,][]{Payne1994,Salas2017}. The line profile due to Doppler broadening is a Gaussian function while the profiles due to pressure and radiation broadening are Lorentzian functions. The net line profile is a Voigt profile that results from the convolution of the Gaussian and Lorentzian profiles. The FWHM line width of the Voigt profile can be expressed as
\be{\Delta V \approx 0.53\Delta V_L + \sqrt{0.22\Delta V_L^2 + \Delta V_D^2},}
\ee
where $\Delta V_D$ and $\Delta V_L$ are the FWHM widths of the Gaussian and Lorentzian profiles respectively. 
For a given $T_e$ and background temperature $T_c$, we can compute the Voigt line width $\Delta V$ as a function of frequency and determine the $n_e$ values that are consistent with the observed data. { An implicit assumption in the modeling is that a homogeneous gas with one LSR velocity component is present along the sightline. If multiple line forming regions of similar temperature but with different LSR velocities are present, then the derived electron density should be considered as an upper limit. }

Careful estimation of the line width is thus crucial to constrain the physical parameters of the line forming region. We model the line profiles at 750 and 321 MHz with Gaussian functions, as the `Lorentzian wings' are not apparent at these frequencies. A Voigt profile is necessary to model the 150 MHz line feature. The spectral baseline removal of the 150 MHz spectrum is critical as it affects the measured line width. Line parameters obtained after removing a 3rd order polynomial from the 150 MHz spectrum and those obtained without baseline removal are given in Table~\ref{tab:rrlpar} (see Fig.~\ref{fig:3crrls}). The line parameters listed are the amplitude, central velocity and the FWHM Lorentzian line width. We fix the value of $\Delta V_D$ as the width of the line at 750 MHz for the Voigt profile modeling as the total contribution from radiation and pressure broadening at this frequency is only about 0.5\% of the observed line width.

\begin{figure*}
\begin{center}
\includegraphics[width=0.45\textwidth]{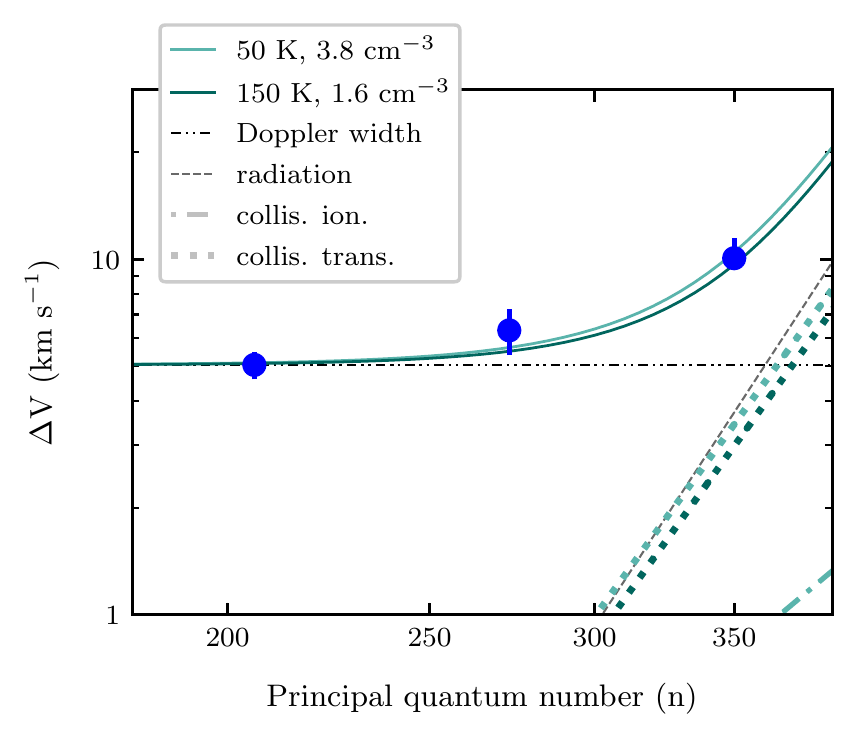}
\includegraphics[width=0.54\textwidth]{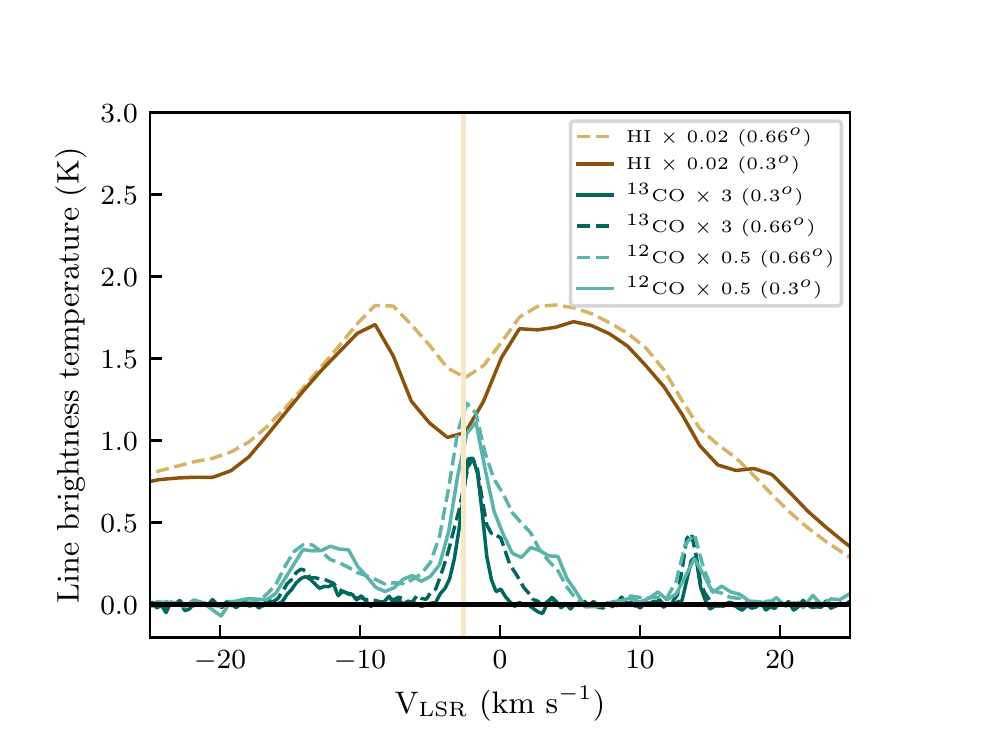}
\caption{\textbf{Left:} FWHM line widths of CRRLs observed toward DR4 vs principal quantum number. The observed points along with $\pm$ 1$\sigma$ error bars are shown for 321 and 750 MHz observations. For 150 MHz, the effect on the line width due to baseline removal from the spectrum is also included in the error bar (see text). 
The solid curves represent the expected line broadening as a function of quantum number for two different models (see Section~\ref{subsec:dr4crrl}), and also shown are the broadening contributions from Doppler motions (dash dot dotted), radiation broadening with $\alpha=2.67$ (dashed), collisional ionizations (dot dashed) and collisional transitions (dotted) relevant for each model.
\textbf{Right:} \HI\ , $^{12}$CO and $^{13}$CO lines toward DR4. The spectra obtained by averaging over 0.66\ddeg\ and 0.3\ddeg\ are shown in dashed and solid lines respectively. {The vertical line indicates the LSR velocity of the CRRL detected at 750 MHz.} 
}
\label{fig:dr4lw_hico} 
\end{center}
\end{figure*}

The electron densities obtained from the line width modeling are in the range 1.4 $\rightarrow$ 6.5 cm$^{-3}$ for the assumed electron temperature range 200 $\rightarrow$ 20 K. The temperature range is chosen based on the modeling results of low-frequency CRRL observations toward other directions in the Galaxy \citep[eg.][]{Kantharia2001}.
The models used for background temperature are given in Table~\ref{tab:contpar}. The values for $T_c$ at 1430.4 MHz and their uncertainties are obtained from the CGPS image and the spectral index values, $\alpha$, are taken from \citet{Ladouceur2008}. The range of the derived physical parameters includes uncertainties in the line widths and the variation in $T_c$ at different frequencies. We have included the range of line widths derived from the 150 MHz spectra with and without baseline removal in the uncertainty of the line width. Two example plots of the expected line widths as a function of quantum number along with the observed values are shown in Fig.~\ref{fig:dr4lw_hico}(left).  


The LSR velocities of the spectral lines observed toward DR4 are shown in Fig.~\ref{fig:DR4HIall}. While the 750 and 321 MHz CRRLs have similar central velocities (within 1$\sigma$), the central velocity of the 150 MHz CRRL is $\sim$ 0.8 \kms (4$\sigma$) lower than that of the 750 MHz line. 
Based on the similarity of the observed LSR velocities of the \HI\ self-absorption feature and CRRLs at 750 and 321 MHz (see Fig.~\ref{fig:DR4HIall}), we assume the cold \HI\ gas and the CRRL region co-exist.  We can therefore assume that the electron temperature of the CRRL forming region is approximately equal to the kinetic temperature of the cold \HI\ gas, which is, in turn approximately the same as its spin temperature.
For a representative electron temperature $T_e = 50$ K, consistent with the upper limit obtained for $T_s$, the estimated electron densities are in the range 2.8 - 4.5 \cmthree. For these model parameters, the radiation broadening is in the range 3.4 to 4.1 \kms, collisonal transition broadening is in the range 2.5 to 4.1 \kms\ and collisonal ionization broadening is in the range 0.6 to 0.9 \kms\ at 150 MHz.   The neutral density of the region from which CRRLs originate is estimated as $\frac{n_e}{\delta A_c}$ where $A_c = 2.9 \times 10^{-4}$ is the cosmic abundance of carbon and $\delta = 0.48$ is the depletion factor \citep{Jenkins2009}. The estimated neutral density is in the range 2 - 3.2 $\times 10^4$ \cmthree. 

 
\subsubsection{Gas properties derived using CO data}

The $^{12}$CO \citep{Dame2001} and $^{13}$CO \citep{Schneider2006} spectra averaged over the GBT beam are shown in Fig.~\ref{fig:dr4lw_hico} (right). The parameters of the Gaussian line model of the components of interest here are listed in Table~\ref{tab:hicopar}. The central velocities of the CO lines overlap with those of the CRRLs (within 2$\sigma$) but have an offset of $\sim$ 1 \kms\ relative to the LSR velocity of the \HI\ self-absorption feature (see Fig.~\ref{fig:DR4HIall} bottom-left).  The $^{12}$CO/$^{13}$CO integrated line ratios obtained from 0.3$^{o}$ and 0.66$^{o}$ apertures
are 12 and 4 respectively. The excitation temperature for $^{12}$CO, estimated from the 0.3\ddeg\ averaged spectrum and by assuming that the $^{12}$CO line is optically thick at its peak, is $\sim$ 5 K. The low values for line ratio and excitation temperature indicate that the gas volume density is modest. 

We use the results presented by \citet{Goldsmith2008} (see also \citealt{Pineda2010}; their Table 1 and Fig. 3) from models developed with a large velocity gradient approximation to get an estimate of the $^{12}$CO column density $N_{CO}$ and H$_2$ density. The modeling also assumes a spherical cloud with uniform kinetic temperature of 15 K and the levels are subthermally excited (i.e. not in LTE). The model results were obtained for $^{12}$CO/$^{13}$CO abundance ratios range between 25 and 65, as this ratio is expected to vary due to chemical and/or photo effects. Their model results provide $N_{CO} \sim 10^{16}$ cm$^{-2}$ and $n_{H_2} \sim 200$ cm$^{-3}$.  For such low density, diffuse clouds, the $H_2$ column density per K \kms\ of $^{12}$CO line emission is 2 $\times 10^{20}$ cm$^{-2}$/(K \kms) \citep{Liszt2010}. Using this relationship, we find the H$_2$ column density of the CO emitting cloud toward DR4 is $\sim 10^{21}$ cm$^{-2}$. The range in the above derived quantities is a factor of 1.5, which is determined from the values estimated using the line parameters from the 0.3\ddeg\ and 0.66\ddeg\ averaged spectra.

\subsubsection{Physical picture}

The neutral density derived from the CRRL line width is the total hydrogen density i.e. $n_H + 2\; n_{H2}$, where $n_H$ is the density of the atomic hydrogen. The total hydrogen density is in the range 2 - 3.2 $\times 10^4$ \cmthree, which is more than 50 times the density ($n_{H_2} \sim$ 200 \cmthree) derived from CO data. The thermal pressure in the CRRL forming region is $nT_e \sim 1 - 1.6 \times 10^6$ \cmthree\ K, where $n$ is the estimated neutral density of the region. The thermal pressure of the CO emitting gas is $\sim$ 3000 \cmthree\ K, at least 300 times smaller than the CRRL forming region. A likely scenario is that the CRRL forming region and the cold \HI\ gas reside in a shocked region at the boundary of, or within the diffuse CO emitting gas. The SNe which produced the remnant DR4 could be a source of the shock.

\subsection{DR21}
\label{sec:dr21}

The compact \HII\ region DR21 is part of a large complex of radio sources located in the Cygnus X region at a distance of 1.5 kpc \citep{Rygl2012}. { Radio and infrared recombination line observations made toward DR21 have provided information about the ionization, the spatial and velocity structures of the ionized gas and the properties of the interface region between the \HII\ region and the neutral cloud in its vicinity \citep{Pankonin1977, Vallee1987, Roelfsema1989, Golynkin1991}. Low angular resolution ($>$ 1$^{'}$) 
observations have detected a change in the central velocity of the hydrogen RRL from $-$5 to 6 \kms\ over the frequency range 8.6 to 1.4 GHz \citep{Pankonin1977}. This change in velocity is interpreted in terms of a simple model involving compact \HII\ regions surrounded by an envelope of lower density ionized gas. The \HII\ regions are ionized by at least 6 OB stars as inferred from interferometric RRL observations \citep{Roelfsema1989}}. Line emission from many molecular species, as well as \CI\ and \CII\ FIR lines have { also } been observed in this direction \citep[e.g.,][]{Jakob2007}
at central velocities $\sim$ -10 to +20 \kms. The bright molecular line emissions are concentrated in the velocity range $\sim$ -10 to 0 \kms\ and $\sim$ +8 to +9 \kms. These features are 
associated with two interacting giant molecular clouds. The negative velocity cloud is part of the DR21 complex and the +8 to +9 \kms\ velocity cloud is part of the W75 complex \citep{Dickel1978, Gottschalk2012}. Cold \HI\ gas observed as \HI\ self-absorption features was also reported in this region near the two velocities \citep{Gottschalk2012}.  

\subsubsection{Carbon and Heavy element RRLs}
\label{subsec:dr21crrl}

Fig.~\ref{fig:3crrls} shows the CRRL spectra observed toward DR21 at 750 and 321 MHz as well as the LOFAR spectrum at 150 MHz. Three line components are detected near the expected velocity range of the carbon line in the 750 MHz spectrum. These components have LSR velocities with respect to the rest frequency of carbon of -11.5, -2.1 and 8.6 \kms. Careful examination of the 321 MHz GBT spectrum indicates that there are line emission features 
near velocities -2.1 and 8.6 \kms. Examination of the 150 MHz spectrum clearly shows an absorption line near -2.1 \kms, and subtracting a single component Voigt profile fit from the spectrum shows some excess absorption near 8.6 \kms\ (see Section~\ref{subsec:dr21crrlwidth} for further discussion on Voigt fit to the 150 MHz spectrum). However, no emission or absorption features are seen in the 321 and 150 MHz spectra near -11.5 \kms. It is likely that this line at 750 MHz may be a transition due to a higher mass element. 
For the analysis in the subsequent sections, we consider only two carbon line components at velocities -2.1 and 8.6 \kms.

\begin{figure*}
\begin{center}
\includegraphics[width=0.45\textwidth]{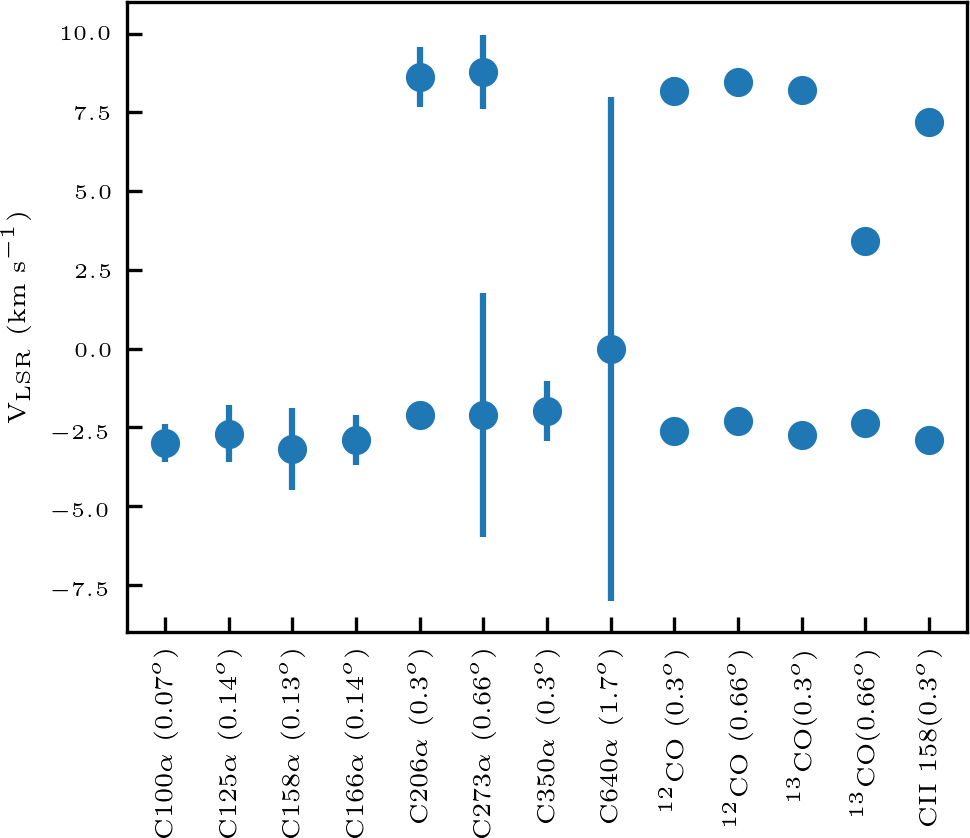}
\includegraphics[width=0.54\textwidth]{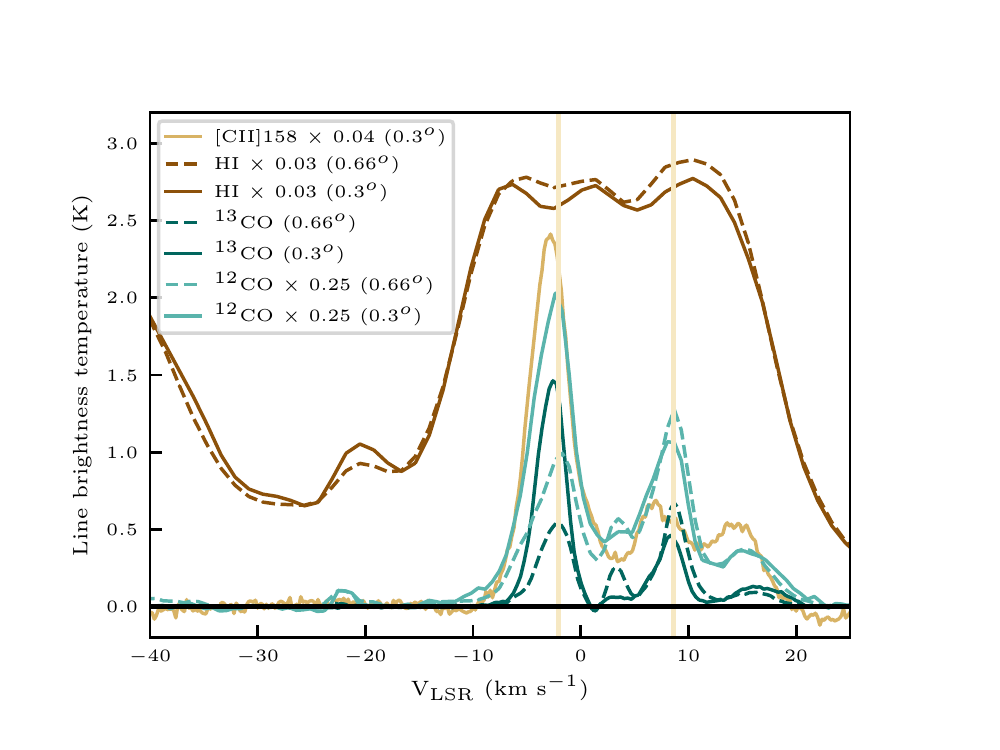}
\caption{\textbf{Left:} {The LSR velocities of CRRLs, $^{12}$CO, $^{13}$CO and CII 158 $\mu$m lines observed toward DR21. The error bars shown are $\pm$3$\sigma$ values for all data points except for the C640$\alpha$ line. The error bar for the C640$\alpha$ line is $\pm 1 \sigma$. For the CO and CII 158 $\mu$m lines the error bars are within the size of the marker.}  
\textbf{Right:} \HI\, $^{12}$CO, $^{13}$CO and CII 158 $\mu$m  spectra toward DR21. The angular resolution and scaling applied to the spectra are indicated in the legend. The vertical lines indicate the LSR velocities (-2.1 and 8.6 \kms) of the CRRLs detected at 750 MHz.} 
\label{fig:dr21hicocii_lsrv} 
\end{center}
\end{figure*}

{  
We list the parameters of the previous CRRL detections \citep{Pankonin1977, Vallee1987, Golynkin1991} in Table~\ref{tab:otherrrlpar} in order to compare with those obtained in our observations. A plot of the LSR velocities of the CRRLs, $^{12}$CO, $^{13}$CO and CII 158 $\mu$m lines (see Section~\ref{sec:dr21hico} and~\ref{sec:dr21cii158}) observed in the direction of DR21 is shown in Fig.~\ref{fig:dr21hicocii_lsrv}(left). CRRLs with positive ($\sim$ 8.6 \kms) LSR velocity are observed only at frequencies $<$ 1.4 GHz.  The uncertainty of the central velocity of the absorption feature detected at 25 MHz \citep{Golynkin1991} is large (about 8 \kms) compared to the errors in the central velocities of other observations; the absorption feature is likely to coexist with the negative velocity CRRL component detected in our observations \citep[see also][]{Golynkin1991}. The mean value of CRRL emission lines detected at frequencies $>$ 1.4 GHz is -3.0$\pm$0.3 \kms, consistent (within 3$\sigma$ value) with the -2.1 \kms\ component observed in our observations, which indicates that they may be associated.  CO and CII 158$\mu$m components corresponding to both CRRL velocity features were detected in the spectra obtained with 0.3\ddeg\ aperture (see Fig~\ref{fig:dr21hicocii_lsrv} right).}   

\subsubsection{ Constraints on the gas properties from CRRL width} 
\label{subsec:dr21crrlwidth}

We follow the method described in Section~\ref{subsec:dr4crrl} for modelling the CRRL line width as a function of frequency to constrain the gas properties toward DR21. The parameters of the Gaussian line models obtained from the 750 and 321 MHz spectra are given in Table~\ref{tab:rrlpar} (see Fig.~\ref{fig:3crrls}). The signal-to-noise ratio of the -2.1 \kms\ component in the 321 MHz spectrum is low, which results in larger fractional errors in the estimated line parameters than were found in DR4. We attempt to fit a two component Voigt profile to the 150 MHz spectrum, using both velocity centroids from the higher frequency data, but the estimated error in the line parameters is comparable to the derived values. Therefore, the 150 MHz spectrum is modeled using a single component Voigt profile. The total contribution to the line width at 750 MHz due to radiation and pressure broadening is less than 5\% of the observed line width at this frequency.  {The estimated error of the line width at 750 MHz is 0.1 \kms, a factor of 2 better than the higher frequency measurements from the literature, and the angular resolution at this frequency is comparable (within a factor of 2) to the low frequency CRRL observations.} Hence for the Voigt profile fitting we used $V_D = 4.1$ \kms, the observed line width at 750 MHz. We carefully examined the spectral baseline and subtracted a 3$^{rd}$ order polynomial before modeling the line profile. The line parameters are given in Table~\ref{tab:rrlpar}. We also included line parameters that were obtained from the spectrum without any baseline removal. The line widths for the two cases differ by a factor of 1.7 (see Table~\ref{tab:rrlpar}). Thus the effect of spectral baseline removal on the line width is substantial for the DR21 direction.  

\begin{figure*}
\begin{center}
\includegraphics[width=0.54\textwidth]{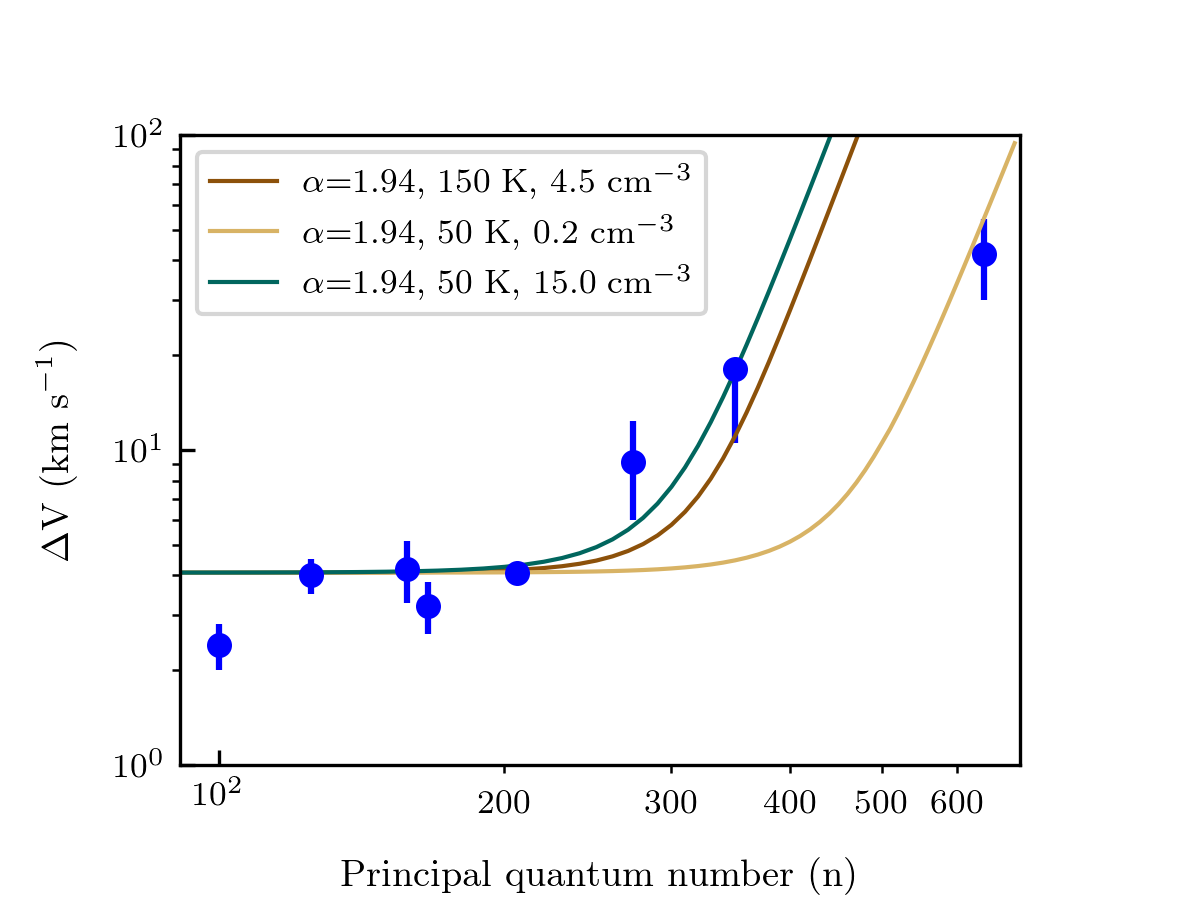}
\includegraphics[width=0.45\textwidth]{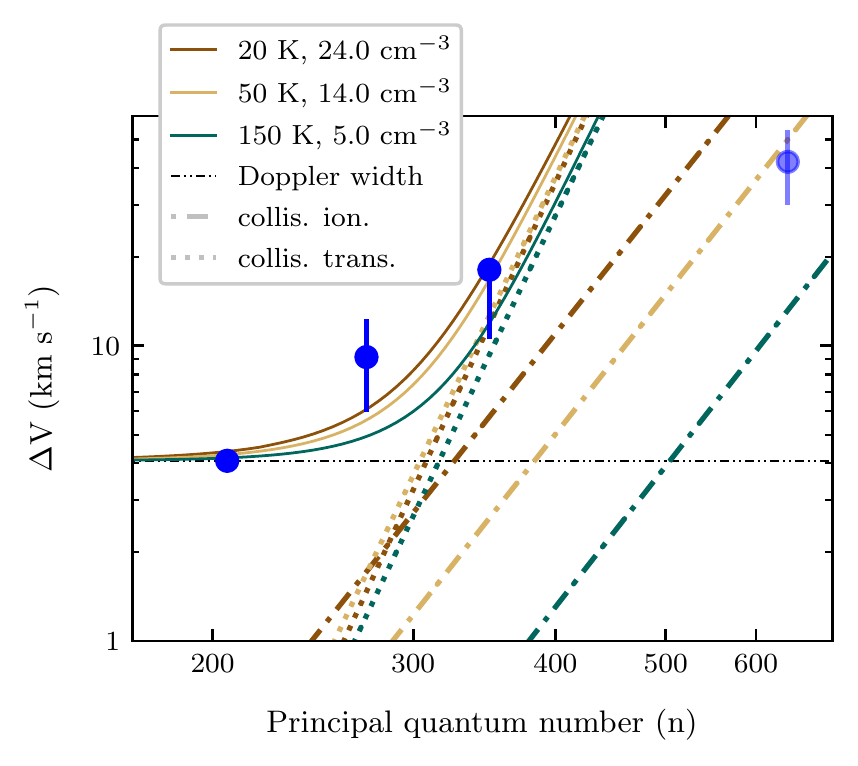}
\caption{\textbf{Left:} {The FWHM line widths of CRRLs observed toward DR21 vs principal quantum number. The $\pm$ 1$\sigma$ error bars are shown for all observed data points except for the 150 MHz line width. The error bar of the 150 MHz line width includes the range of values obtained with and without spectral baseline removal. 
The solid curves represent the expected line broadening as a function of quantum number for three different models. It is evident from the figure that at least two CRRL forming regions with very different electron densities are required to explain the observed line width as a function of frequency (see Section~\ref{subsec:dr21crrlwidth}).  
\textbf{Right:} Same as the left figure but showing observed points at frequencies $<$ 1.4 GHz. The solid curves represent three different models for the higher density line forming region (see Section~\ref{subsec:dr21crrlwidth}), and also shown are the broadening contributions from Doppler motions (dash dot dotted), collisional ionizations (dot dashed) and collisional transitions (dotted) relevant for each model. Radiation broadening modeled with an $\alpha=1.94$ does not significantly contribute to line broadening in this region.}} 
\label{fig:dr21lw} 
\end{center}
\end{figure*}

{We plot the observed line width of the -2.1 \kms\ CRRL component as a function of quantum number in Fig.~\ref{fig:dr21lw}(left). Three example models for line width variation with quantum number are shown in the figure.}
{The line width for the C$100\alpha$ line observed at an angular resolution of $4.3\arcmin$ \citep{Vallee1987} is not consistent with the those of the lower frequency CRRLs. This difference can be attributed to the different angular resolution between the observations, as well as changes in the gas probed as a function of frequency.}
{As seen in Fig.~\ref{fig:dr21lw}(left), for a given $T_e =$ 50 K, the electron densities required to explain the C350$\alpha$ (150 MHz) and the C640$\alpha$ (25 MHz) absorption lines \citep{Golynkin1991} differ by a factor of 75. The situation is similar for other electron temperatures as well. Thus, at least two line forming regions are needed to explain the observed line widths as a function of frequency. Below we attempt to separately constrain the physical parameters of these regions. }

{We first consider the gas responsible for the 150 MHz absorption line. Fig.~\ref{fig:dr21lw}(right) shows results from the line width modeling along with observed values at frequencies $<$ 1.4 GHz. We could not find models that are consistent with the 321 MHz observations for $T_e$ \gsim 50 K when a 1$\sigma$ range is considered for the observed values. An example model result that is consistent with all the observed line widths between 750 and 150 MHz requires $T_e =$ 20 K, $n_e =$ 24 \cmthree\ (see Fig.~\ref{fig:dr21lw} right). Since the signal-to-noise ratio of the -2.1 \kms\ component at 321 MHz (C$273\alpha$) is low, we include here models that are consistent only with the 750 and 150 MHz observations. The estimated electron densities from the line width modeling are in the range  3.5 $\rightarrow$ 24  cm$^{-3}$ for assumed electron temperatures in the range 200 $\rightarrow$ 20 K. The range of neutral densities corresponding to the estimated electron densities is 2.5 - 17.2 $\times 10^{4}$ \cmthree. We refer to this gas as the high density CRRL forming region.} 

{We now consider the gas responsible for the 25 MHz CRRL absorption. The line width modeling indicates that the electron density of the gas responsible for the 25 MHz line absorption is in the range 0.008 $\rightarrow$ 0.3 \cmthree\ for the assumed range of electron temperature 200 $\rightarrow$ 20 K. Example model curves are shown in Fig.~\ref{fig:dr21lw}(left). The estimated parameter values are consistent with the constraints on the gas properties obtained by \citet{Golynkin1991}. The neutral density of the line forming region obtained from the estimated electron densities is in the range 58 - 2200 \cmthree. We refer to this gas as the low density CRRL forming region. 
}

{
\subsubsection{Gas properties using CRRL amplitude: preliminary results} 
\label{subsec:dr21crrlamp}

We take representative values for the gas parameters from the range of possible values estimated in Section~\ref{subsec:dr21crrlwidth}  for the dual line forming regions: $T_e \sim 150$ K, $n_e \sim 4.5$ \cmthree\ and LOS pathlength 7 $\times 10^{-3}$ pc for the high-density region and $T_e \sim 50$ K, $n_e \sim 0.2$ \cmthree\ and LOS pathlength 0.25 pc for the low density region (see Fig.~\ref{fig:dr21lw} left). These parameter values are not uniquely determined, which will be attempted as part of our future modeling work (see Section~\ref{sec:sum}).
The higher density region could be a photo-dissociation region (PDR) located at the \HII\ region-molecular cloud interface and the lower density region could be located in the diffuse interstellar medium along the LOS toward the source. The background continuum temperature at 750 MHz is $\sim$ 99 K and so the detection of emission lines from both regions at this frequency indicates that the carbon atom level population is not in LTE. Therefore, it is essential to have the non-LTE parameters to compute the expected line intensity from the dual line forming region model; here we use the results from \cite{Walmsley1982}. 

Our modeling shows that the observed CRRL at 150 MHz is dominated by the line absorption due to the high density region as its electron density is high. The CRRL from the lower density region is in emission at 150 MHz and is amplified by stimulated emission due to the background radiation field as its electron density is low. The relative contributions to line emission from the high and low density regions at 321 MHz are 10\% and 90\% and at 750 MHz are 45\% and 55\% respectively. The $b_n \beta_n$ values obtained from \cite{Walmsley1982} for the low density region parameters are -7.2, -7.4 and -2.6 for quantum numbers 206, 273 and 350 respectively. We use the $b_n \beta_n$ values -5.7, -0.6 and 6.0 for quantum numbers 206, 273 and 350 respectively for the high density region, which corresponds to $T_e=100$ K and $n_e=$ 1 \cmthree\ \citep{Walmsley1982} {as departure coefficients for higher temperatures and densities are not available}. We also assumed the dielectronic-subthermal case discussed by \cite{Walmsley1982} for the modeling.} 


\subsubsection{\HI\ and CO lines}   
\label{sec:dr21hico}

The \HI\ , $^{12}$CO \citep{Dame2001} and $^{13}$CO \citep{Schneider2010} spectra toward DR21 averaged over the 0.3\ddeg\ and 0.66\ddeg\ beam are shown in Fig.~\ref{fig:dr21hicocii_lsrv} (right). The \HI\ line shows a self-absorption feature near (within 1$\sigma$) the -2.1 \kms\ CRRL component but no absorption feature is seen near the 8.6 \kms\ component (see Fig.~\ref{fig:dr21hicocii_lsrv} right). The \HI\ line structure, however, is more complex than that toward DR4 and so we could not perform a Gaussian decomposition to extract the line parameters of the absorption feature. The CRRL central velocities are similar (within 1$\sigma$) to the two features seen in the CO spectra at -2.1 and 8.6 \kms. The CO line amplitudes in the 0.66\ddeg\ averaged spectra are lower by a factor of $\sim$2 compared to those in the 0.3\ddeg\ averaged spectra, which indicates that the physical properties of the molecular cloud change over this angular scale. The isotopologue ratio and excitation temperature (assuming $^{12}$CO is optically thick at the line center) derived from the 0.66 deg averaged spectra are 11 and 7 K respectively. From the models of \citet{Goldsmith2008}, the $^{12}$CO column density and H$_2$ densities are 7 $\times 10^{16}$ cm$^{-2}$ and 275 cm$^{-3}$ respectively for an assumed kinetic temperature of 15 K. The estimated H$_2$ column density is 4 $\times 10^{21}$ cm$^{-2}$ \citep{Liszt2010}. 

\subsubsection{Physical picture} 

{
The higher density region responsible for CRRL absorption at 150 MHz has a neutral density more than 100 times the molecular density inferred from CO line emission. The LOS extent of the molecular cloud is about 4.5 pc, at least 2 orders of magnitude larger than that of the dense CRRL forming region. 
The enhancement in density could be due to shocks resulting from the 
expansion of the \HII\ region DR21. Thus it is likely that the dense CRRL forming region is a PDR located at the boundary of the DR21 and the associated molecular cloud. The lower density region responsible for the CRRL absorption at 25 MHz, on the other hand, has neutral density a factor of $\sim$5 times the CO gas density. Its LOS extent is 20 times lower than the CO cloud size. Thus the lower density region could be residing in a denser part of the diffuse CO emitting cloud. The enhancement in density could be due to shocks resulting from the interaction of the two giant molecular clouds on this sight line. }

\subsubsection{CII 158 $\mu$m emission}
\label{sec:dr21cii158}

The CII 158 $\mu$m line was detected toward DR21 by SOFIA with an angular resolution of 0.5\ddeg $\times$ 0.2\ddeg\ \citep[see Fig~\ref{fig:dr21hicocii_lsrv} right;][]{Schneider2020}. The line central velocity of the strongest emission is -2.92 \kms (see Table~\ref{tab:hicopar}). This velocity is 0.8 \kms\ offset from the -2.1 \kms\ CRRL component; about 8 times the uncertainty of the central velocity of the recombination line. The integrated line temperature of this component is 26.4 K \kms. We estimate the expected CII 158 $\mu$m intensity from the dual CRRL forming region using Eq. 2 of \citet{Roshi2002}. {The critical densities for electron, hydrogen atom and hydrogen molecule collisions are assumed to be 10, 3000 and 6000 cm$^{-3}$ respectively for electron temperatures in the range 50 to 150 K. The total hydrogen atom densities used for the calculation are estimated from the derived electron densities for the dual CRRL forming region (see Section~\ref{subsec:dr21crrlamp}). 
We found that the dual line forming region model can account for 73\% of the observed CII 158 $\mu$m line intensity, with 58 \% and 14\% contributions from the high and low density CRRL forming regions respectively. }


\subsection{HB21}
\label{sec:hb21}

HB21 is an evolved, nearby SNR of age 8000-15000 yrs, with an angular extent of $\sim$ 2.5\ddeg $\times$ 2.3\ddeg \citep{Kothes2006}. $^{12}$CO, \HI\ and radio continuum observations indicate that the SNR is interacting with the interstellar material \citep{Tatematsu1990}. The 21 cm line observations show an HI shell around the SNR expanding at $\sim$ 25 \kms \citep{Assousa1973}. A giant molecular cloud east of the remnant has been observed and the CO emission morphology indicates that HB21 is interacting with the molecular cloud \citep{Tatematsu1990}.

\subsubsection{Gas properties derived using \HI\ data}

\begin{figure*}
\begin{center}
\plottwo{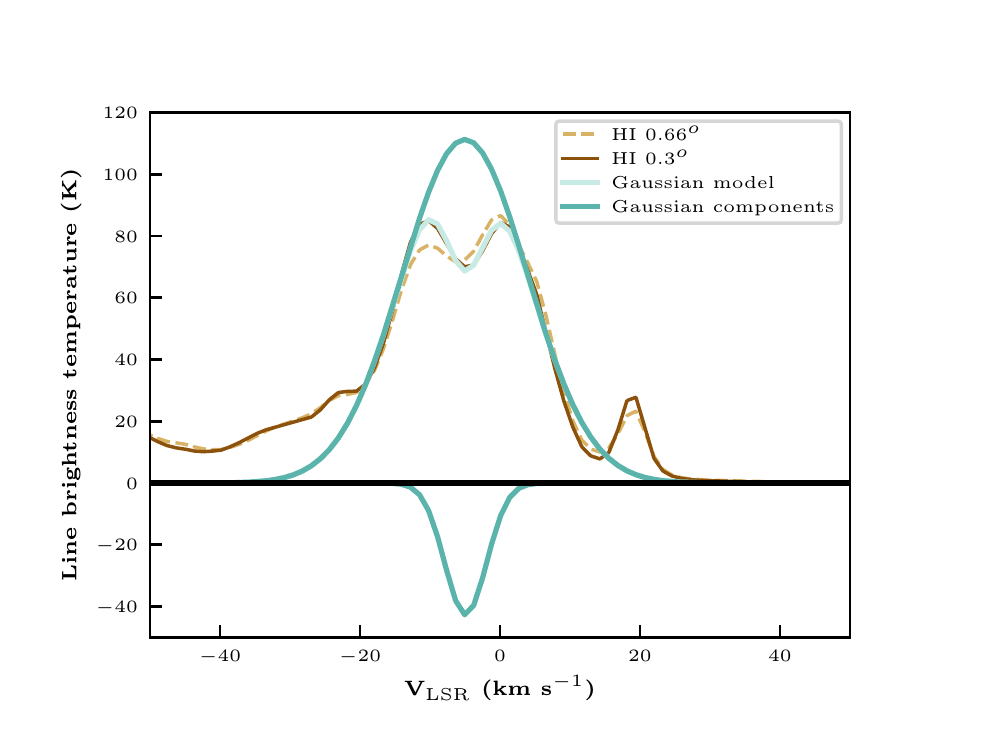}{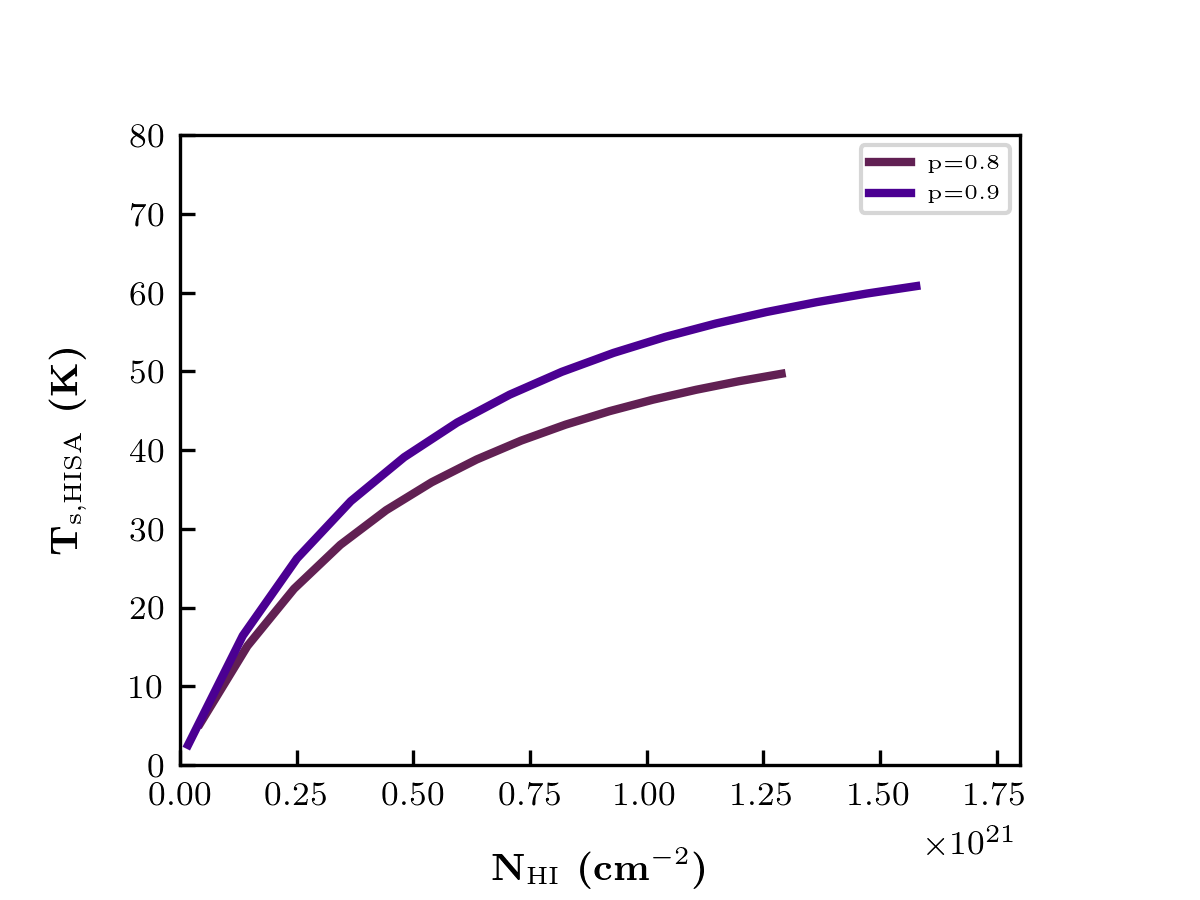}
\caption{\textbf{Left}: \HI\ 21cm spectra toward HB21 along with Gaussian line model of the line profile. The angular resolution is indicated in the legend. \textbf{Right}: {Constraints} on the physical parameters of the cold \HI\ gas in the direction of HB21 (see Section~\ref{sec:hb21}).
}
\label{fig:hb21hisa} 
\end{center}
\end{figure*}

The \HI\ spectra averaged over the 0.3\ddeg\ and 0.66\ddeg\ regions are shown in Fig.~\ref{fig:hb21hisa}.  The \HI\ line near -5 \kms\ shows a strong emission superposed with a self-absorption feature. As in the case for DR21 and DR4, the \HI\ profile is complex and a unique Gaussian decomposition could not be done. We modeled the self-absorption feature using a single Gaussian emission component and an absorption component. For the line profile modeling the velocity range was restricted to -20 to 80 \kms. The result of the Gaussian line modeling is shown in Fig.~\ref{fig:hb21hisa} and the \HI\ line parameters are given in Table~\ref{tab:hicopar}. The properties of the cold gas responsible for \HI\ self-absorption are constrained using the method described in Section~\ref{subsec:dr4hi}. Fig~\ref{fig:hb21hisa} shows the \HI\ spin temperatures and column densities of the cold gas that can produce the observed self-absorption feature. The \HI\ column densities required to produce the observed absorption feature are at least 50 \% larger than those of the cold gas observed toward DR4.

\subsubsection{Constraints from CRRL observations}

Our 750 MHz observations did not detect CRRL emission towards HB21. The upper limit on the line amplitude is 5 mK (see Table~\ref{tab:rrlpar}). Assuming an electron temperature of 50 K and taking $b_n \beta_n \sim -3.6$  \citep{Walmsley1982} at 750 MHz, we get an upper limit for the emission measure of 9 $\times 10^{-3}$ pc cm$^{-6}$. This emission measure is {comparable to that of the lower density CRRL forming region towards DR21.} 

\subsection{G34.20+0.0, G34.94+0.0 \& G35.17+0.0}

CRRL emission is detected towards G34.94+0.0 and tentatively detected toward G34.20+0.0 at 327 MHz (see Fig.~\ref{fig:spec2}). Diffuse CRRL emission and absorption have been previously detected in the inner galactic plane at frequencies above and below $\sim$ 100 MHz respectively \citep{Erickson1995,Kantharia2001,Roshi2002}. The CRRLs detected with the Arecibo telescope are likely to be from diffuse \CII\ regions similar to those observed in earlier carbon line surveys. The line width of the CRRL toward G34.94+0.0 is 18.5 \kms, much larger (at least a factor of 4) than seen towards DR21 and DR4 at 321 MHz, but closer to the typical line width detected in an earlier 327 MHz survey of CRRLs \citep{Roshi2002}. The larger line width could be from contributions to the observed emission from multiple line forming regions along the LOS with different LSR velocities.  
Future multi-frequency observations will be necessary to model the line forming region and understand its origin.

\subsection{Hydrogen and Helium RRLs}

Hydrogen lines are detected toward almost all observed positions except in the direction of HB21. We estimate the emission measure (EM) of the \HII\ regions responsible for the hydrogen line emission assuming, LTE, electron temperature of 8000 K and a beam filling factor of unity using the equation
\be{
EM = \frac{\int T_L d\nu}{1.92 \times 10^3\; T_e^{1.5}},
}\ee
where EM is in units of pc cm$^{-6}$, $T_e$ is in K and $\int T_L d\nu$ is the observed integrated line temperature in K kHz.
The emission measures obtained from the 321 and 750 MHz data are 8100 and 1.1 $\times 10^4$ pc cm$^{-6}$ respectively toward DR4. The emission measures estimated toward DR21 are 8500 and 1.5 $\times 10^4$ pc cm$^{-6}$ from the observed lines at the two frequencies. We attribute the {difference in EMs obtained from the two frequencies} to beam dilution effects reflecting clumpiness in the ionized gas. 

The emission measure estimated toward G34.94+0.0 is 5500 pc cm$^{-6}$. We examined the WISE catalogue \citep{anderson2014} to identify \HII\ regions in this area. The \HII\ region G034.940+0.074 \citep{Lockman1989} is present within the Arecibo telescope beam (15$^{'}$),
but its LSR velocity is 45.6 \kms; no strong hydrogen RRL component at 327 MHz was detected at that velocity. A second \HII\ region, G034.757-0.669, has LSR velocity 52.1 \kms, which agrees with the central velocity of the hydrogen RRL observed toward G34.94+0.0 at 327 MHz, but is at a separation of 0.7\ddeg\ from the sight line.

The emission measure estimated from the detected hydrogen line toward G34.20+0.0 is $\sim$ 6000 pc cm$^{-6}$.  There are two \HII\ regions - G034.256+00.136, G034.404+00.227 - located respectively at angular distances of 0.15\ddeg\ and 0.3\ddeg\ from G34.20+0.0, with LSR velocities 53.1 and 60.1 \kms. The LSR velocity of G034.404+00.227 is within the 1$\sigma$ uncertainty of the central velocity of hydrogen line detected toward G34.20+0.0. 

Toward G35.17+0.0, we estimate an emission measure of $\sim$ 3000 pc cm$^{-6}$. The \HII\ region G035.063+00.330 (angular distance 0.35\ddeg ) has LSR velocity 57.2 \kms\ similar to the central velocity of the detected line toward G35.17+0.0. 

We conclude that the Hydrogen RRLs in the Galactic Plane likely arise in 
the outer envelopes of \HII\ regions. \citep{Anantharamaiah1986}. 

A Helium line is detected towards DR21 at 750 MHz. The {helium to hydrogen} line ratio is 0.08 (0.02). Earlier observations at frequencies of 1.4 GHz and above have found that this ratio varies from 0.03 to 0.95 within the DR21 region \citep{Roelfsema1989}. However these observations were made with an angular resolution at least a factor of 2 higher than the 750 MHz observations reported here.

\section{Summary and future work}
\label{sec:sum}

Radio recombination lines of carbon are detected toward DR4 and DR21 at 750 and 321 MHz with the GBT. These observations are combined with 150 MHz LOFAR {and other previous} CRRL detections to constrain the physical properties of the line forming regions. {Modeling the line width as a function of frequency indicates that the electron density of the line forming region toward DR4 is in the range 1.4 $\rightarrow$ 6.5 cm$^{-3}$. Similar modeling shows that a dual CRRL forming region with electron densities 3.5 $\rightarrow$ 24 cm$^{-3}$ and 0.008 $\rightarrow$ 0.3 \cmthree\ could plausibly explain the observed line width as a function of frequency toward DR21. The electron densities of the denser regions in both directions} are at least a factor of 10 larger than those estimated for diffuse \CII\ regions in the Galactic plane from low ($<$ 1 GHz) frequency observations \citep[e.g.,][]{Kantharia2001,Oonk2017,Salas2017}. $^{12}$CO and $^{13}$CO emission and \HI\ self-absorption lines at similar central velocities to the CRRLs are detected toward DR4 and DR21. {The CII 158 $\mu$m lines observed toward DR21 also have central velocities similar to those of the CRRLs.} The similarity of the central velocities suggests that these line forming regions are for the most part associated.  The inferred molecular densities from the CO data are at least two orders of magnitude smaller than that of the {denser CRRL forming regions in both directions}. We suggest that the CRRLs and \HI\ self-absorption are formed in shocked regions at the boundary of or within the cloud producing the CO emission {in the direction of DR4. The shocks, which caused the enhancement of the density, could be due to SNe blast waves in the case of DR4. The denser line forming region toward DR21 could be located at the \HII\ region-molecular cloud interface and the density enhancement could be due to shocks caused by the expansion of the \HII\ region. The lower density CRRL forming region toward DR21 has a neutral density about 5 times the density of the CO cloud observed in this direction. This density enhancement could be due to shock caused by cloud-cloud collision. Our modeling shows that a significant fraction ($\sim$ 73\%) of the observed CII 158 $\mu$m emission toward DR21 could originate from the dual CRRL forming region.}

We also detect CRRLs toward G34.20+0.0 (tentatively) and G34.94+00 near 327 MHz with the Arecibo telescope. These CRRLs could be from diffuse \CII\ regions similar to those detected earlier in the Galactic plane \citep{Anantharamaiah1986}.

Hydrogen lines were detected in almost all directions except towards HB21, and at all the observed frequencies. A Helium line was detected toward DR21 at 750 MHz and the observed helium to hydrogen line ratio is 0.08 (0.02).

We plan to expand the GBT observations over a larger area in the Cygnus region to probe the spatial extent of the CRRL forming region and to better understand its association with regions producing CO and \HI\ self-absorption lines. We also plan to make multi-frequency observations toward the Galactic plane positions where CRRLs were detected with the Arecibo telescope. Our next step in modeling the line emission is to develop a time dependent photodissociation model which can self-consistently model the \HI\, CO and CRRLs. This model will also be able to correctly calculate the level population of CRRLs at high Rydberg states. The larger data set and 
temporal information will help to better constrain the physical properties of the absorbing clouds.  The eventual goal is to derive the molecular formation rates, thus addressing the wider question of molecular cloud formation.  

\section{Acknowledgements} \label{sec:acknowledgements}
The Arecibo Observatory is a facility of the NSF operated under cooperative agreement (\#AST-1744119) by the University of Central Florida (UCF) in alliance with Universidad Ana G. M\'{e}ndez (UAGM) and Yang Enterprises (YEI), Inc. The Green Bank Observatory is a major facility funded by the National Science Foundation operated by Associated Universities, Inc. This publication makes use of data products from the Wide-field Infrared Survey Explorer, which is a joint project of the University of California, Los Angeles, and the Jet Propulsion Laboratory/California Institute of Technology, funded by the National Aeronautics and Space Administration. This research has made use of NASA’s Astrophysics Data System. Phil Perillat, Arecibo Observatory, was crucial in helping us with the observational setup and initial observations. Basic Research at the Naval Research Laboratory is funded by 6.1 base programs. We acknowledge the very productive discussion with F. J. Lockman, Green Bank Observatory, during the course of this work. {JBRO acknowledges financial support from NWO Top LOFAR-CRRL project, project No. 614.001.351. LOFAR, the Low Frequency Array designed and constructed by ASTRON, has facilities in several countries, that are owned by various parties (each with their own funding sources), and that are collectively operated by the International LOFAR Telescope (ILT) foundation under a joint scientific policy. The authors would like to thank the LOFAR observatory staff for their assistance in obtaining and handling of this large data set. Part of this work was carried out on the Dutch national e-infrastructure with the support of the SURF Cooperative through grant e-infra 160022 \& 160152. We thank the anonymous referee whose very helpful comments have significantly improved the manuscript.}  

\facility{Arecibo Telescope, GBT, {LOFAR}}

\bibliographystyle{aasjournal}
\bibliography{crrl327.bib}{}

\begin{deluxetable*}{lccrcccc}[b!]
\tablecaption{Summary of Observations \label{tab:obssum}}
\tablecolumns{7}
\tablewidth{0pt}
\tablehead{
\colhead{Source} &
\colhead{RA(2000)} &
\colhead{DEC(2000)} &
\colhead{Date of Obs.} & \colhead{Obs. freq} & \colhead{Beam} & \colhead{Telescope} \\
\colhead{Name} & \colhead{(hh:mm:ss)} & \colhead{(dd:mm:ss)} &
\colhead{yyyy-mm-dd} & \colhead{(MHz)} & \colhead{(\ddeg )} & \colhead{}
}
\startdata
DR4  (G78.12+1.92) & 20:21:56    & +40:15:36  & 2012-05-09              & 321,750 & {0.66, 0.3} & GBT  \\
DR21 (G81.68+0.54) & 20:39:01    & +42:19:43  & 2012-05-09, 2012-05-27  & 321,750 & {0.66, 0.3} & GBT \\
HB21 (G89.00+4.70) & 20:46:05    & +50:39:05  & 2012-05-09              & 750     & 0.66 & GBT \\
G34.20+0.0         & 18:53:45.1  & +01:07:43  & 2019-06-03, 2020-05-11  & 327     & 0.25     & Arecibo \\
G34.94+0.0         & 18:55:06.1  & +01:47:14  & 2019-06-02, 2020-05-08 & 327     & 0.25     & Arecibo \\
                   &             &            & 2020-06-01, 2020-06-02  &         &          &  \\
G35.17+0.0         & 18:55:31.3  & +01:59:30  & 2019-05-27, 2020-04-26  & 327     & 0.25     & Arecibo \\
\enddata
\end{deluxetable*}

\begin{deluxetable*}{cccrrrr}[b!]
\tablecaption{Observed RRL parameters \label{tab:rrlpar}}
\tablecolumns{7}
\tablewidth{0pt}
\tablehead{
\colhead{Freq} & \colhead{Atom} & \colhead{n} & \colhead{T$_{LA}$} & \colhead{V$_{LSR}$} & \colhead{$\Delta$V} & \colhead{Note} \\
\colhead{(MHz)}  & \colhead{ }  & \colhead{ }         & \colhead{(K)}         & \colhead{\kms}  & \colhead{(\kms)} & \colhead{}
}
\startdata
\multicolumn{7}{c}{DR4  (G78.12+1.92)}\\\hline
321 & H   & 273 & 0.77(0.05) &  1.4(0.8) & 26.5(1.9) &  \\
    & C   &     & 0.77(0.09) & -2.8(0.4) & 6.6(0.9) & \\
750 & H   & 206 & 0.46(0.01) &  0.9(0.2) & 25.0(0.5)  &  \\  
    & C   &     & 0.24(0.02) &  -2.6(0.2)  &  5.1(0.5)  &  \\
150 & C   & 350 &  -4.2(0.1) &  -1.7(0.1) & $\Delta V_L =$7.4(0.9) & \tablenotemark{1} \\
    & C   &     & -3.8(0.1) &  -1.8(0.1) & $\Delta V_L =$8.5(1.0) & \tablenotemark{2} \\
    & C   &     & -4.3(0.2) &   -1.7(0.2) & $\Delta V_L =$7.5(1.3) & \tablenotemark{3} \\
    & C   &     & -3.8(0.1) &   -1.8(0.2) & $\Delta V_L =$7.3(1.2) & \tablenotemark{4} \\
    \hline
\multicolumn{7}{c}{DR21  (G81.68+0.54)}\\\hline
321 & H   & 273 & 0.76(0.03) &  7.2(0.5) & 28.1(1.3)  &  \\
    & C   &     & 0.63(0.05) & 8.8(0.4) & 7.0(0.9) & \\
    & C   &     & 0.22(0.05) & -2.1(1.3)& 9.3(3.2) &  \\
750 & H   & 206 & 0.66(0.01) & 6.9(0.1) & 25.3(0.2)&  \\
    & He  &     & 0.05(0.01) & 7.8(1.0) & 16.2(2.4)&  \\
    & C   &     & 0.21(0.01) & -2.1(0.1) & 4.1(0.3)& \\
    & C   &     & 0.11(0.01) & 8.6(0.3) & 7.4(0.8) & \\
    & S   &     & 0.09(0.02) & -2.9(0.2)& 2.4(0.5) & \tablenotemark{10} \\
150 & C   & 350 & -1.1(0.1)  & -2.0(0.3) & $\Delta V_L =$17.1(2.1) & \tablenotemark{5} \\ 
    & C   &     & -1.1(0.1)  & -1.6(0.3) & $\Delta V_L =$17.2(1.9) & \tablenotemark{6} \\
    & C   &     & -1.1(0.1)  & -2.3(0.3) & $\Delta V_L =$10.2(2.3) & \tablenotemark{7} \\
    & C   &     & -1.0(0.1)  & -1.9(0.3) & $\Delta V_L =$10.6(2.2) & \tablenotemark{8} \\
\hline    
\multicolumn{7}{c}{HB21 (G89.00+4.70)}\\\hline
750 &     & 206 & (0.005)    &               &              & \\\hline
\multicolumn{7}{c}{G34.20+0.0}\\\hline
327 &  H  & 272 & 0.26(0.03) & 62.8(3.6) & 58.6(8.6)  & \tablenotemark{9}\\
    &  C  &     & 0.25(0.06) & 49.8(2.1)    & 11.0(4.9)   &\tablenotemark{9} \\\hline
\multicolumn{7}{c}{G34.94+0.0}\\\hline
327 & H   & 272 & 0.43(0.04) &  52.1(1.6) & 32.7(3.7)  & \\
    & C   &     & 0.24(0.05) &  49.4(2.2)  & 18.5(5.1)  & \\\hline
\multicolumn{7}{c}{G35.17+0.0}\\\hline 
327 &  H  & 272 & 0.15(0.01) & 56.9(2.7) & 48.5(6.4) & \tablenotemark{9}\\
\enddata
\tablenotetext{1}{Parameters obtained from the spectrum averaged over 0.3\ddeg . A 3$^{rd}$ order polynomial baseline was removed from the spectrum. $\Delta V_D =$5.1 \kms}
\tablenotetext{2}{Parameters obtained from the spectrum averaged over 0.66\ddeg . A 3$^{rd}$ order polynomial baseline was removed from the spectrum. $\Delta V_D =$5.1 \kms}
\tablenotetext{3}{Parameters obtained from the spectrum averaged over 0.3\ddeg . $\Delta V_D =$5.1 \kms}
\tablenotetext{4}{Parameters obtained from the spectrum averaged over 0.66\ddeg . $\Delta V_D =$5.1 \kms}
\tablenotetext{5}{Parameters obtained from the spectrum averaged over 0.3\ddeg . A 3$^{rd}$ order polynomial baseline was removed from the spectrum. $\Delta V_D =$4.1 \kms}
\tablenotetext{6}{Parameters obtained from the spectrum averaged over 0.66\ddeg . A 3$^{rd}$ order polynomial baseline was removed from the spectrum. $\Delta V_D =$4.1 \kms}
\tablenotetext{7}{Parameters obtained from the spectrum averaged over 0.3\ddeg . $\Delta V_D =$4.1 \kms}
\tablenotetext{8}{Parameters obtained from the spectrum averaged over 0.66\ddeg . $\Delta V_D =$4.1 \kms}
\tablenotetext{9}{Tentative detection}
\tablenotetext{10}{see Section~\ref{sec:dr21}}
\end{deluxetable*}

\begin{deluxetable*}{lccr}[b!]
\tablecaption{\HI\ , CO and CII 158$\mu$m line parameters \label{tab:hicopar}}
\tablecolumns{4}
\tablewidth{0pt}
\tablehead{
\colhead{Amplitude} & \colhead{Center} & \colhead{Width} &\colhead{Note} \\
\colhead{(K)}         & \colhead{(\kms)}   & \colhead{(\kms)} & \colhead{}
}
\startdata
\multicolumn{4}{c}{DR4  (G78.12+1.92): \HI\ }\\\hline
   2.2(1.0) &  48.4(2.4)  &   10.9(5.7) & \tablenotemark{1} \\
  99.0(1.2) &   -1.2(0.1) &   33.0(0.4) & $T_{OFF}$ \\
 -48.0(1.5) &   -3.2(0.1) &    6.8(0.3)&  $T_{ON} - T_{OFF}$ \\
  32.9(1.0) &  -33.4(0.4) &  13.6(1.0)  &  \\
  33.3(1.4) &  -47.0(0.3) &   9.5(0.7)  &  \\
  28.7(2.9) &  -62.5(0.3) &   11.3(1.0) &  \\ 
  65.7(0.8) &  -78.3(0.5) &   27.2(0.8) &  \\
\hline
\multicolumn{4}{c}{DR4  (G78.12+1.92): $^{12}$CO }\\\hline
  2.1(0.1) & -2.0(0.1) &  3.2(0.1) & \tablenotemark{1}\\
  1.3(0.1) & -2.4(0.1) &  2.2(0.1) & \tablenotemark{2}\\
\hline
\multicolumn{4}{c}{DR4  (G78.12+1.92): $^{13}$CO }\\\hline
  0.30(0.01) &   -2.05(0.02) & 1.86(0.04) & \tablenotemark{1} \\
  0.28(0.01) &   -2.08(0.06) & 2.72(0.12) & \tablenotemark{2}\\
\hline
\multicolumn{4}{c}{DR21  (G81.68+0.54):$^{12}$CO }\\\hline
 7.7(0.1) & -2.6(0.1) & 5.0(0.1) & \tablenotemark{1} \\
 4.3(0.2) &  8.2(0.2) & 4.0(0.4) & \tablenotemark{1} \\
 3.7(0.1) & -2.3(0.1) & 5.7(0.2) & \tablenotemark{2} \\
 4.9(0.1) &  8.5(0.1) & 3.9(0.2) & \tablenotemark{2} \\
\hline
\multicolumn{4}{c}{DR21  (G81.68+0.54):$^{13}$CO }\\\hline
  1.47(0.01) & -2.75(0.01) & 3.15(0.03) & \tablenotemark{1}\\
  0.43(0.01) &  8.20(0.05) & 3.26(0.11) & \tablenotemark{1}\\
  0.55(0.01) & -2.38(0.03) & 3.47(0.07) & \tablenotemark{2}\\
  0.63(0.01) &  8.66(0.03) & 3.03(0.06) & \tablenotemark{2}\\
\hline
\multicolumn{4}{c}{DR21  (G81.68+0.54):CII 158$\mu$m }\\\hline
  5.84(0.04) & -2.92(0.02) & 4.52(0.04) & \tablenotemark{3}\\
\hline
\multicolumn{4}{c}{HB21 (G89.00+4.70): \HI\ }\\\hline
 -39.3(4.8) &  -4.8(0.2)  &    7.0(0.7) & \tablenotemark{1}\\
 108.1(4.9) &  -4.9(0.1)  &   22.1(0.8) & \\
\enddata
\tablenotetext{1}{Parameters obtained from the spectrum averaged over 0.3\ddeg .}
\tablenotetext{2}{Parameters obtained from the spectrum averaged over 0.66\ddeg .}
\tablenotetext{3}{Parameters from the spectrum obtained with an angular resolution of 0.5\ddeg $\times$ 0.2\ddeg.}
\end{deluxetable*}

\begin{deluxetable*}{lccrr}[b!]
\tablecaption{Continuum parameters used for modeling \label{tab:contpar}}
\tablecolumns{5}
\tablewidth{0pt}
\tablehead{
\colhead{Source} & \colhead{$f_0$\tablenotemark{a}} & \colhead{$T_C$} &\colhead{$\alpha$} & \colhead{Note} \\
\colhead{}       & \colhead{(MHz)} & \colhead{(K)}   & \colhead{}        & \colhead{} 
}
\startdata
DR4 & 1420.4 & 46(4) & 2.67 (2.63 $\rightarrow$ 2.72) & \tablenotemark{1} \\
    & 1420.4 & 42(4) & 2.67 (2.63 $\rightarrow$ 2.72) & \tablenotemark{2} \\
\hline
DR21& 150.0  & 2246(10) & 1.94(0.01) & \tablenotemark{3}\\
    & 150.0  & 2149(10) & 2.0(0.02) & \tablenotemark{4} \\
\hline
HB21& 1420.4 & 10(4) & 2.38 & \tablenotemark{5} \\
\enddata
\tablenotetext{a}{Frequency corresponding to the listed continuum temperature $T_c$. The continuum temperature at a frequency $f$ in MHz can be obtained as $T = T_C ( \frac{f}{f_0})^{-\alpha}$ }
\tablenotetext{1}{Continuum values obtained with 0.3\ddeg\ beam. The median spectral index from \cite{Ladouceur2008} and
the interquartile range are listed.}
\tablenotetext{2}{Continuum values obtained with 0.66\ddeg\ beam. The median spectral index from \cite{Ladouceur2008} and
the interquartile range are listed.}
\tablenotetext{3}{Continuum values obtained with 0.3\ddeg\ beam. The spectral index was obtained from 408 and 1420 MHz
CGPS measurements.}
\tablenotetext{4}{Continuum values obtained with 0.66\ddeg\ beam. The spectral index was obtained from 408 and 1420 MHz
CGPS measurements.}
\tablenotetext{5}{Continuum values obtained with 0.3\ddeg\ beam. The spectral index was from \cite{Kothes2006}. }
\end{deluxetable*}

\begin{deluxetable*}{cccrrr}[b!]
\tablecaption{CRRL parameters from previous observations toward DR21 \label{tab:otherrrlpar}}
\tablecolumns{6}
\tablewidth{0pt}
\tablehead{
\colhead{Freq} & \colhead{n} & \colhead{T$_{LA}$} & \colhead{V$_{LSR}$} & \colhead{$\Delta$V} & \colhead{Note} \\
\colhead{(MHz)} & \colhead{ }         & \colhead{(K)}         & \colhead{\kms}  & \colhead{(\kms)} & \colhead{}
}
\startdata
1425 &  166 &  0.1(0.02)    &  -2.9(0.3) & 3.2(0.6) & \tablenotemark{1}\\
1652 &  158 &  0.1(0.02)    &  -3.2(0.4) & 4.2(0.9) & \tablenotemark{2}\\
3329 &  125 &  0.017(0.002) &  -2.7(0.3) & 4.0(0.5)  & \tablenotemark{3} \\  
6482 &  100 &  0.011(0.002) &  -3.0(0.2) & 2.4(0.2)  & \tablenotemark{4} \\
25   &  640 &  -35(15)      &   0(8)     & 42(12)    & \tablenotemark{5} \\
\enddata
\tablenotetext{1}{Angular resolution of the observation is 8.5$^{'}$\citep{Pankonin1977}.}
\tablenotetext{2}{Angular resolution of the observation is 7.8$^{'}$\citep{Pankonin1977}.}
\tablenotetext{3}{Angular resolution of the observation is 8.2$^{'}$\citep{Vallee1987}.}
\tablenotetext{4}{Angular resolution of the observation is 4.3$^{'}$\citep{Vallee1987}.}
\tablenotetext{5}{Equivalent beam width of the observation is 1.7$^{o}$\citep{Golynkin1991}.}
\end{deluxetable*}

\end{document}